\documentclass[aps,superscriptaddress,prd,twocolumn,showpacs]{revtex4}
							
\usepackage{amsmath}
\usepackage{subeqnarray}
\usepackage{graphicx}	
\usepackage{epstopdf}      
\usepackage{pdfsync}

\begin{document}

\newcommand{\etal}{\textit{et al}} 
\title{Phenomenological Relations for  Axial Quasi-normal Modes of Neutron
  Stars with Realistic Equations of State}
\author{J. L. Bl\'azquez-Salcedo}
\affiliation{
Depto. F\'{\i}sica Te\'orica II, Facultad de  Ciencias F\' \i sicas,\\
Universidad Complutense de Madrid, 28040-Madrid, Spain}
\author{L. M. Gonz\'alez-Romero}
\affiliation{
Depto. F\'{\i}sica Te\'orica II, Facultad de  Ciencias F\' \i sicas,\\
Universidad Complutense de Madrid, 28040-Madrid, Spain}
\author{F. Navarro-L\'erida}
\affiliation{
Depto. F\'{\i}sica At\'omica, Molecular y Nuclear, Facultad de  Ciencias F\'
\i sicas,\\ 
Universidad Complutense de Madrid, 28040-Madrid, Spain}
\date{\today}
\begin{abstract}
Here we investigate the axial w quasi-normal modes of neutron stars  for
18 realistic equations of  
state, most of them satisfying the $2 M_\odot$ condition. In particular, we study the influence of the presence of hyperons and quarks in
the core of the neutron stars. 
We have obtained that 
w-modes can be used to 
distinguish between neutron stars with
exotic matter and without exotic matter for compact enough stars.
We present phenomenological relations for the frequency and damping times with
the compactness of the  
neutron star for wI and wII modes showing the differences of the stars with
exotic matter  in the core. 
Also, we obtain a new phenomenological relation between the real part and the
imaginary part of the frequency  
of the w quasi-normal modes, which can be used to estimate the central
pressure of the neutron stars. Finally, we study the low compactness limit configuration of
fundamental wII modes, and the influence of
changes in the core-\textit{crust} transition pressure.   
To obtain these results we have developed a new method based on the Exterior
Complex Scaling technique with variable angle.
\end{abstract}
\pacs{04.40.Dg, 04.30.-w, 95.30.Sf, 97.60.Jd}
\maketitle
\section{Introduction}
Considerable progress has been made in the last years on the development of 
gravitational wave detectors. Large-scale interferometric
gravitational wave detectors, like LIGO, GEO, TAMA and VIRGO, have reached the
original design sensitivity, and are currently beginning to operate with
frequencies between 1Hz to 1kHz.  The sensitivity is continuously being
enhanced \cite{GW_Living_Review2011}.  
First detections are expected to happen within the next five years. These
observations are of huge importance 
because they can be used to perform stringent tests of General Relativity. But
also because gravitational wave detection opens a new 
window to observations of the insights of numerous, and probably also new,
astrophysical processes. Possible sources of gravitational waves are
interacting black holes, coalescing compact binary systems, stellar collapses
and pulsars \cite{GW_Living_Review2009}. 

Neutron stars are major candidates to detectable gravitational wave sources
because of 
their rich emission spectrum, which lays inside the frequency range of current
detectors. A recent  review about gravitational waves from neutron stars can be found in  \cite{Andersson_etal}. The study of the signature of the equation of state on gravitational radiation from neutron star has been developed in \cite{Andersson_obs_1996, Shibata_etal, 
Shibata_2, Janka,Hinderer_etal, Read_etal}. The first direct search for the gravitational-wave emission associated with oscillations of fundamental quadrupole mode excited by a pulsar glitch has been presented in \cite{Abadie_etal} . 

Neutron stars are found inside pulsars, and are thought to originate after the 
collapse of massive star cores. The supernova explosion causes violent
oscillations of the resulting compact star. Similarly the coalescence
of compact bodies like white dwarfs leaves behind an oscillating neutron
star. The resulting excess of energy that causes the star to oscillate is 
radiated in the form 
of gravitational radiation \cite{GW_Living_Review2009}.   

It is well known that although the spectrum of neutron star oscillations is
continuous, general perturbations causes the star to ring with concrete
oscillation frequencies that dominates over the rest of possible
frequencies for a certain period of time after the perturbation. These
resonant frequencies can be studied introducing the concept 
of quasi-normal modes, that is, eigen-modes of oscillation, that, although do
not form a complete set of functions in which to expand every possible
perturbation evolution, are very useful in the determination of the
eigen-frequencies for which the star tends to oscillate
\cite{Kokkotas_Schmidt1999,Nollert1999,Rezzolla2003ua}. These eigen-frequencies
are given by a 
complex number. The real part gives us the oscillation frequency of the mode,
while the 
imaginary part gives us the inverse of the damping time. The quasi-normal mode
spectrum is quite dependent on the properties of the star, i.e. the equation
of state.  
 
Neutron stars are compact objects of very high density. Inside of them matter
is found at extreme densities ($10^{15} g/cm^{3}$). Current theories
predict a layer structure for the neutron star, essentially composed by the
core and the 
\textit{crust}. The properties of the \textit{crust} are very different from
those of the core, and it is thought that this region has a solid crystalline
structure similar to a metal. Although the high density matter inside the
star has exotic properties, the 
resulting relativistic fluid that compose the neutron star can be described as
a perfect fluid. What we need to know is the equation of state for the
neutron star matter, but beyond the nuclear densities this relation is not
well understood. The inner part of the
core is very model dependent, essentially because different populations of
particle states may appear at those densities. The composition is not exactly
known and several hypothesis exist
\cite{haensel2007neutron,glendenning2000compact}. Along the core of the star
and specially in the core-\textit{crust} interface, 
first order phase transitions are expected to be found in 
realistic equations of state. These phase transitions result in small
discontinuities in 
the energy density of the star matter
\cite{haensel2007neutron,heiselberg2000}. 

The measurement of $1.97 M_\odot$ for PSR J1614-2230 imposes a stringent condition on the equations of state, in particular to those 
with exotic matter in the core \cite{Lattimer_Prakash}. Recently several
equations of state with exotic matter in the core satisfying the $2 M_\odot$
condition have been proposed \cite{Bednarek,Sedrakian,Weissen1,Weissen2}.

 In order to systematize the study of constraints placed by astrophysical
 observations on the nature of neutron star matter, several 
 parameterizations of high-density equations of state (EOS) have been introduced: a piece-wise polytropic
approximation by Jocelyn et al. \cite{Jocelyn2009} and  Lindblom spectral decomposition
\cite{Lindblom2010}. The advantage of these
parameterizations is that  they 
reduce the total amounts of parameters which modelize the equation of state to
a more tractable quantity, allowing to constrain with observational data the values of the parameters, and also
usually they enhance a numerical integration scheme
allowing 
better precision. Let us note
that a very large family of equations of state can be fitted to these parameterizations with a very good precision. From a practical/numerical
point of view, another possibility to describe the equation of state,
satisfying  local thermodynamic conditions, is a monotone piecewise cubic
Hermite  interpolation. 
 
The detection of gravitational radiation from neutron stars can be a very
useful tool to 
determine the neutron star equation of state. The extraction of the spectrum
of quasi-normal modes (frequency and damping time) from these
signals could give important constrains to the equation of state, and in
consequence, 
information about the behavior of matter at densities beyond nuclear matter
\cite{Benhar_Berti_Ferrari_1999,Kokkotas2001,Benhar_Ferrari_Gualtieri_2004,Benhar2005,Ferrari2007,Chatterjee2009,Wen2009}.       

The theoretical study of the quasi-normal mode spectrum considering different
models of neutron star with different equations of state is then
well justified. The necessary formalism was developed first for quasi-normal
modes of black holes by Regge and Wheeler \cite{ReggeI} and by Zerilli
\cite{PhysRevLett.24.737}. Quasi-normal modes can be differentiated into polar
and axial modes. In these papers it is found that the equation
describing the quasi-normal mode perturbation of the Schwarzschild metric is
essentially a Schr\"odinger like equation: the Regge-Wheeler equation for axial
perturbations and the Zerilli equation for polar ones. For black holes, both
types of modes are space-time modes. The formalism was 
studied in the context of neutron stars first by Thorne
\cite{ThorneI,ThorneII,ThorneIII,ThorneIV,ThorneV}, Lindblom
\cite{Lindblom1983,DetweillerLindblom1985}, 
and then reformulated by Chandrasekhar and Ferrari 
\cite{Chandrasekhar08021991,Chandrasekhar09091991,Chandrasekhar08081991}, and
Kojima \cite{Kojima1992}. In neutron stars, axial modes are purely 
space-time modes of oscillation (w modes), while polar modes can be coupled
to fluid oscillations (although a branch of w modes can be found also in
polar oscillations). In this paper we will consider only axial modes of
oscillation.

The study of the quasi-normal modes spectrum is complicated because of several  
reasons. Quasi-normal modes can only be studied numerically. No
analytical 
solutions are known for physically acceptable configurations of neutron
stars. Another reason is that quasi-normal modes are found as isolated points
scattered on the complex plane, so that usually an exhaustive scan of this plane
is necessary in order to find the complete spectrum. Also the
numerical study is hindered by the very 
definition of  
quasi-normal mode itself, that as we will see explicitly in the following
sections, gives rise to diverging functions that oscillate an infinite number
of times towards spatial infinity. These functions are
not well handled numerically. 

Several methods have been developed to deal with these
and other difficulties. For a
complete review on the methods see the review by Kokkotas and Schmidt
\cite{Kokkotas_Schmidt1999}. 

Chandrasekhar 
and Ferrari
\cite{Chandrasekhar08021991,Chandrasekhar09091991,Chandrasekhar08081991} used 
a slowly damped approximation ($\Im[\omega]\ll\Re[\omega]$), which reduces
the system of equations needed to resolve the problem, so that the damping
time of the quasi-normal mode can be given in terms of the real part. This
method was also used in the determination of w-modes on polar perturbations of
constant density neutron stars by Kojima et al \cite{Kojima1995}. It is only valid for slowly
damped modes, which are present in highly 
compact stars, and is not appropriate for realistic configurations, although
it can be used to obtain a first approximation of the fundamental
frequencies. 

In   an interesting paper \cite{Kokkotas1994}, Kokkotas studied the axial spectrum integrating the
exterior solution up to a finite radius, which allows to
impose the outgoing wave behavior to the exterior solution. The
matching 
between the interior and exterior solutions results in the construction of a
Wronskian at the 
matching surface, which
must be zero when both solutions correspond to a quasi-normal
mode. 

The WKB method
\cite{Kokkotas1991} can be used to approximate the outer
solutions provided that no back-scattering is found in the exterior
region. This condition  
 usually is satisfied  when the imaginary part of the mode is small
compared with the real part \cite{Andersson1995}. 

The phase
function can be studied in order to deal with the oscillatory nature of the
perturbation function. Studied for black holes by Chandrasekhar and Detweiller
\cite{Chandrasekhar1975}, the phase function plays a fundamental role in the
numerical approach made by Andersson et al in \cite{Andersson1995} and in
\cite{Andersson1996} for neutron stars. The phase is usually a well
behaved function if there is not back-scattering contamination in the
numerical solution.

The divergence of the outgoing wave solution of the perturbation equations
makes the avoidance of incoming wave contamination a difficult task. This
problem was 
dealt for black holes by Andersson in \cite{Andersson1992} by rotating the
radial variable into a complex variable parallel to the anti-Stokes line. This
approach was used for constant density neutron stars in \cite{Andersson1995}
and 
\cite{Andersson1996} together with the phase function approach. More
recently, Samuelsson et al \cite{Samuelsson2007} used a similar complex-radius
approach for a constant
density configuration which dealt with the 
divergence at infinity by integrating the Bondi-Sachs phase along a fixed path
for the complexified radius parallel to the anti-Stokes line.

Another approach is based on the Leaver continued fraction
method \cite{Leaver1985}, which allows to impose the outgoing wave behavior on
the border of the star 
as a self-consistent equation that can be satisfied iteratively
\cite{Leins1993}. This method has been used successfully with realistic models
of neutron stars
\cite{Benhar_Berti_Ferrari_1999,Benhar_Ferrari_Gualtieri_2004,Wen2009} 

As commented above, for fluid modes there is also a branch of
w-modes. As these are purely space-time modes, they do not couple to matter,
and the inverse Cowling approximation can be used to obtain this part of the
spectrum \cite{Andersson1996}.

In this work
we present a new approach to calculate quasi-normal modes of realistic neutron
stars. We make use of several well-known techniques, like the use of the phase
for the exterior solution and the use of a complexified coordinate to deal with
the divergence of the outgoing wave. We also introduce some new techniques not
used before in this context: freedom in the angle of the exterior complex path
of integration, 
implementation of  realistic equations of state in a piece-wise polytrope
approximation or by monotone piecewise cubic Hermite interpolation, use of Colsys package to integrate all the system of equations at once
with proper boundary and junction conditions and possibility of implementation
of phase transition discontinuities. These new techniques  allow us to  enhance
precision,  to obtain more modes in shorter times, and also to
study  several realistic equations of state, comparing results for different
compositions.

In section II we will start presenting the well-known general formalism of
quasi-normal modes. In section III we will present an study of the
junction conditions for a general matching that can include first
order phase transitions or surface energy layers. In most of the results of
the paper we will make use of the  
junction conditions without phase transitions or energy layers.
In section IV we describe the numerical method we have used to obtain our
results, testing the algorithm to obtain already known modes for a simple
model. 
In sections V, VI, and VII we present our results, focusing on 18 different equations of
state:  for pure nuclear matter (Sly, APR4), mixed hyperon-nuclear matter(GNH3, H1, H4, BGN1H1,WCS1, WCS2, BHZBM),  
hybrid stars  (ALF2, ALF4,WSPHS3), hybrid stars with hyperons and quark
color-superconductivity (BS1, BS2, BS3, BS4) and quark stars (WSPHS1, WSPHS2). In section VIII we 
finish the paper with a summary of the main points of the work.
\section{Quasi-normal mode formalism}
In order to fix the notation we will present a brief review of the  formalism
used to describe  axial 
quasi-normal modes. In the following we  use geometrized units
($c=1,G=1$). 

We consider a static spherical space-time with metric
$ds^{2}=e^{2\nu}dt^{2}-e^{2\lambda}dr^{2} - r^{2}(d\theta^{2} 
+ \sin^{2}\theta d\varphi^{2})$. 
The matter inside of
the star is considered perfect fluid with stress-energy tensor
$T^{\mu \nu } = \left( p + \rho \right)u^{\mu}u^{\nu} - p g^{\mu \nu}$, where
$p$ is the pressure, $\rho$ is the energy density and $u$ is the 4-velocity.
At zero order  the equations obtained are well
known. Inside the star we have:
\begin{eqnarray}
\frac{d m}{d r}= 4\pi r^{2}\rho, \label{eq_m}\\
\frac{d p}{d r} = -(\rho + p)\frac{m + 4\pi r^{3} p}{(r-2m)r}, \label{eq_p}\\
\frac{d \nu}{d r} = -\frac{1}{\rho + p}\frac{d p}{d r}, \label{eq_nu} 
\end{eqnarray}
where we must provide a equation of state. Outside we have the Schwarzschild solution with gravitational mass M.  

Following the original papers 
\cite{ThorneI}-\cite{Kojima1992}, we
make perturbations over 
this metric, expanding the perturbation in tensor harmonics, and taking into
account only the axial perturbations. 
It can be demonstrated that the
axial perturbations must satisfy the well-known Regge-Wheeler equation
\cite{PhysRevLett.24.737}. 
\begin{equation}
\frac{d^{2}Z^{lk}}{dr^{2}_{*}}+\left[\omega^{2} - V(r)
\right]Z^{lk}=0, \label{eq_Z} 
\end{equation}
where $l$ and $k$ are the spherical harmonic indexes, $r_{*}$ is the
tortoise coordinate: 
\begin{equation}
r_{*}=\int_{0}^{r} e^{\lambda - \nu}dr,
\end{equation}
and the eigen-frequency of the axial mode is a complex number $\omega =
\omega_{\Re} + i\omega_{\Im}$.
The potential
can be written as:
\begin{equation}
V(r) =  \frac{e^{2\nu}}{r^{3}}\left[l(l+1)r + 4\pi r^{3}(\rho+p)- 6m \right],
\end{equation}
In this work we will consider only
the $l=2$ case.


Axial oscillations do not modify the energy
density or the pressure of the fluid. Hence, for axial modes,
inside the star  only the perturbations on the 4-velocity of the matter
have to be taken into account (Lagrangian displacement).

Note that in general the perturbation function $Z$ is a complex
function. Hence, we have a system of two real second order differential
equations (one for the real part $Z_{\Re}$, and another for the imaginary part
$Z_{\Im}$).

$Z$ has to satisfy  a set of boundary conditions that
can be obtained from the following two requirements
\cite{Chandrasekhar08021991}: i) the perturbation must be regular at the
center of the star, and ii) the resulting quasi-normal mode must be a pure
outgoing wave. 


The second requirement imposes the quasi-normal modes to behave as purely
outgoing waves at radial infinity. 
In general a quasi-normal mode will be
a composition of incoming and outgoing waves, i.e.
\begin{equation}
\lim_{r_{*}\to\infty}Z^{in} \sim e^{i\omega r_{*}}, \lim_{r_{*}\to\infty}Z^{out} \sim e^{-i\omega r_{*}}. \label{asymp_out}
\end{equation}
Note that, while the real part of $\omega$ determines the oscillation frequency
of the wave, the imaginary part of the eigen-value determines the asymptotic
behavior of the quasi-normal mode: 
If we call $\tau=1/\omega_{I}$,
the ingoing and outgoing modes will behave, respectively, as:
\begin{equation}
\lim_{r_{*}\to\infty}Z^{in} \sim e^{-r_{*}/\tau},  \lim_{r_{*}\to\infty}Z^{out} \sim e^{r_{*}/\tau},
\end{equation}
Outgoing quasi-normal modes are divergent
at radial infinity, while ingoing ones tend exponentially to zero as the
radius grows. 

In the next sections we will present our numerical method, which is able to deal with this problem. 
 \section{Junction conditions}
To determine completely the problem, we must study the junction conditions
between the interior solution and the exterior 
solution  on the boundary of the star.
The junction conditions between two space-times have been considered in
classical works by Darmois \cite{Darmois}, Lichnerowicz \cite{Lichnerowicz},
O'Brien and Synge \cite{Synge}, and Israel \cite{Israel}
, and more recently in \cite{Mars_Seno,Vera,Mars}.
In the context of relativistic rotating stars, it has been  considered in
 \cite{Gonzalez-Romero}, \cite{MacCallum_etal} and \cite{Sotani2001}.

 We will impose the usual junction conditions. But, in order to have the possibility to 
study the influence of 
small  changes in the core-\textit{crust} transition
pressure, we will include  the more general case in which the star is surrounded
by a surface layer of energy density. 

 We use Darmois conditions, the intrinsic formulation of these junction conditions, which imposes certain constrains on the continuity of  the fundamental forms of the matching hypersurface $S$ 
 \cite{misner1973gravitation}.

 This surface $S$  is
defined by the points where the 
pressure is null, or more generally, constant. As we have seen
previously the axial oscillations
do not modify the pressure or the density, so the surfaces of constant
pressure of 
the perturbed star are essentially the same as the surfaces of constant
pressure of the static star.
.

Let us consider a surface $S$ of constant pressure inside the star, and a  surface layer on top of $S$ described by a 
surface stress-energy tensor  of  a perfect fluid of the following form: 
\begin{equation}
T_{S}^{\mu \nu }(R)=\varepsilon u_{S}^{\mu }(R)u_{S}^{\nu }(R),
\end{equation}
where,  $S$ is defined by the points where $r=const.=R$,  $\varepsilon$ is the surface energy density of the fluid moving  in $S$, and $u_{S}^{\mu }(R)$ is its  
velocity.  
The continuity of the first fundamental form gives us  two conditions: the
continuity of $\nu$
\begin{equation}
\nu_{int} = \nu_{ext}, \label{nono}
\end{equation}
and the continuity of perturbation function $Z$:
\begin{equation}
Z^{lk}_{int} = Z^{lk}_{ext}. \label{cond_Z}
\end{equation}

The second  fundamental form conditions impose a jump in the pressure if
$\varepsilon \neq 0$. Let the pressure 
in the inner part of the surface $S$, be $p_{int}$, and in the outer part be $p_{ext}$ with $p_{ext} < p_{int}$.

Making the same steps as with the first fundamental form, we obtain the
following conditions for the zero order functions:
\begin{eqnarray}
M_{ext}=M_{int}+4\pi R^{2}\sqrt{1-\frac{2M_{int}}{R}}\varepsilon -8\pi
^{2}R^{3}\varepsilon ^{2},  \label{cond_mass}
\\
\frac{M_{ext}+4\pi
R^{3}p_{ext}}{R^{2}\sqrt{1-\frac{2M_{ext}}{R}}}-\frac{M_{int}+4\pi
R^{3}p_{int}}{R^{2}\sqrt{1-\frac{2M_{int}}{R}}} = 4\pi \varepsilon. \label{cond_pressure}
\end{eqnarray}
In this work we will consider $p_{ext}=0$. That is, the surface of star
with pressure $p=p_{int}$ is being matched directly to the empty exterior
surface of the star, in which case, $M_{ext}$ is just the Schwarzschild mass
M.  If we choose $p_{int}$ as the core-\textit{crust}
transition pressure, this matching condition is equivalent to approximating
exterior 
layers of the star to a surface layer energy density that envelops
the core of the star (surface \textit{crust} approximation).   
The equation (\ref{cond_mass}) can be interpreted as giving the total mass of
the star to zero order ($M_{ext}$), as the addition of three terms, the first
one representing the core mass ($M_{int}\equiv 4 \pi \int_0^R \rho r^2 d
r$), 
the second one representing the mass of the \textit{crust} and the third one
being a negative bounding energy term. 
The radius of the resulting configuration will be the radius of the core of
the star. Equation (\ref{cond_pressure}) can be seen as a condition for the
determination of the radius of the core of the star when approximating the
 \textit{crust} of the star to a thin surface energy density.

In the surface \textit{crust} approximation, these relations allow us to use the core-\textit{crust} transition pressure $p_{int}$,  
as a new parameter of the resulting configuration. 

Up to first order in the perturbative expansion we obtain junction conditions
for the axial functions. Essentially we obtain the following condition for the
derivative of the $Z$ function:
\begin{equation}
\frac{dZ^{lk}}{dr}|_{ext} =
\frac{e^{\lambda_{ext}}}{e^{\lambda_{int}}}\frac{dZ^{lk}}{dr}|_{int} +
\left[\frac{e^{\lambda_{ext}}}{e^{\lambda_{int}}}-1\right]\frac{Z^{lk}}{R}. \label{cond_dZ} 
\end{equation}
Together with condition (\ref{cond_Z}) we have enough information to perform
correctly the matching of the perturbation. 

Let us note that,  the usual treatment without surface \textit{crust}
approximation is recovered    by making in the previous
conditions $p_{int}=p_{ext}=0$ and $\varepsilon=0$, in which case the continuity of the metric
functions and of their derivatives is obtained. This is what we do in almost all the paper 
except in the final part of section VII, where the influence of the changes in the core-\textit{crust}
transition pressure is analyzed.
\section{Numerical method}
In this section we will explain the numerical method we use to obtain
outgoing quasi-normal modes of realistic stars. We implement this method into
different Fortran programs and routines, making use of the Colsys package
\cite{colsys1979} to solve numerically the 
differential equations. The advantage of this package is that it allows the
utilization of quite flexible multi-boundary conditions, that as we will see in the
following sections, are important in our method of quasi-normal mode
determination. On the other hand, Colsys allows to construct an adaptative
mesh of points depending 
of the precision we would like to achieve. In this manner, the more
precision we need on the perturbation function, the more points Colsys
introduces into the mesh. The static solution is calculated
at the same time as the perturbation, so the static solution functions will
be also 
adapted to the precision we require to the Regge-Wheeler
function. Colsys also allows the use of the continuation method: we can use as
an initial guess for a new numerical solution another numerical solution which
parameters are close enough to the parameters of the new solution.

We will start
explaining how the outgoing modes are determined in the exterior region of the
metric. 

Quasi-normal modes present an
oscillating part given by the real part of the eigen-frequency, that remains
towards infinity, as can be seen in relation (\ref{asymp_out}). 

In order to deal with this issue we will follow
\cite{Andersson1992}-\cite{Samuelsson2007}, and we will study the phase function.  
The phase function (logarithmic derivative of the Z function),
\begin{equation}
g = \frac{Z'}{Z},
\end{equation}
does not oscillate towards asymptotic infinity. 

The other important issue we have seen is that 
quasi-normal modes diverge towards spatial infinity because of their pure
outgoing wave behavior. This problem can be treated using the phase function together  
with a rotation of the 
radial coordinate into the 
complex plane. 
 

This technique of
complexification of the integration variable is called Exterior Complex   
Scaling \cite{aguilar_combes_1971,balslev_combes_1971,simon1972}. 
An analytical extension of 
the $g$ function, which is a function of $r$, is made into a $g$ function that is
dependent 
of the generalized complex coordinate.
This new function has no physical
meaning, but the eigen-values obtained by integrating the equation along the
extended radius will remain unaltered. 

We parameterize the complex coordinate along
a line in the complex plane. We will take a straight line with an angle
$\alpha$ with respect the real axis.
\begin{equation}
r = R + ye^{-i\alpha},
\end{equation}
with $y\in [0,\infty)$. 
The g function has the same limits in the case
of a purely outgoing or a purely ingoing wave than with the real
coordinate. But the mixed waves now depend 
on the 
angle of the path of integration $\alpha$. 

In our method, the angle of the
integration path is treated as a free parameter of the solution. This angle
can be chosen appropriately to increase precision and 
integration time, but always remaining in the region where $g \to
-i\omega$. 
This boundary condition impose to the obtained solution a purely outgoing wave behavior.  
Exterior Complex Scaling with variable angle 
is widely used in other contexts like atomic and molecular physics, where it
gives 
excellent results 
\cite{Alvarez1991}. 

The use of the phase and Exterior Complex Scaling with variable angle allows
us to compactify, so 
we can impose the outgoing
quasi-normal mode behavior as a 
boundary condition at infinity without using any cutoff for the radial
coordinate.  

Inside the star the Regge-Wheeler equation (\ref{eq_Z}) can be solved without any
problem. The perturbation function is oscillating, but the number of
oscillations will be finite between $r=0$ and $r=R$.


We must provide an equation of state.
We implement the realistic EOS in two different ways:  1) A piece-wise polytrope
approximation, done by Read et al \cite{Jocelyn2009}. 
In this approximation
the equation  of state is approximated by a polytrope in different
density-pressure intervals. For densities below the nuclear density, the SLy
equation of state is considered. 
2) A piecewise monotone 
cubic Hermite interpolation satisfying local thermodynamic condition \cite{Fritsch-Carlson}.

 

We generate two independent solutions of the Regge-Wheeler equation inside the star.
The junction conditions (\ref{cond_Z}) and (\ref{cond_dZ}) can be written in the form of a determinant making use of these two solutions together with the exterior phase function. 
The determinant can be calculated numerically from the integrated
solutions. 



We must explore the plane $(\omega_{\Re},\omega_{\Im})$ looking for null
determinants. 
This process can be automated so that the program looks for minima of the
determinant on the plane. 

We have made several tests on our method using simpler equations of
state. We successfully reproduce data from previous works. For example in table
\ref{tab:test} we show our results for a star of constant density and
$2M/R = 0.885$, 
in perfect agreement with \cite{Samuelsson2007}.

\begin{figure}
\includegraphics[angle=-90,width=0.4\textwidth]{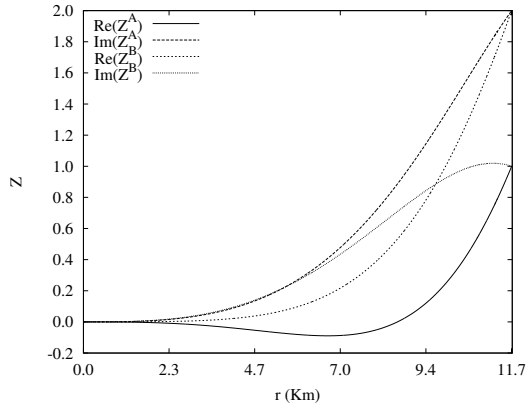}
\caption{Example of Regge-Wheeler function inside the star vs radius.}
\label{example_1}
\end{figure}

\begin{figure}
\includegraphics[angle=-90,width=0.4\textwidth]{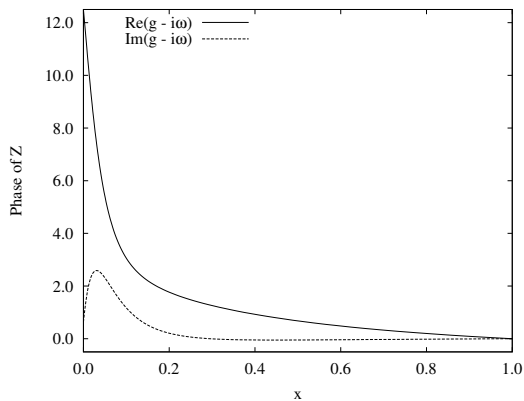}
\caption{Example of phase
  function g outside the star vs the compactified rotated coordinate x.}
\label{example_2}
\end{figure}

In Figures \ref{example_1} and \ref{example_2}
we show an example of
calculated perturbation functions. These functions correspond to a star of $1.4
M_{\odot}$, with Sly equation of state. The quasi-normal mode
is the first axial mode $w_{I}$ with frequency $\nu=8.034 kHz$ and damping
time $\tau =
29.31\mu s$. 
In Figure \ref{example_1} we plot the real and imaginary parts  of the
two interior solutions for Z vs r. In Figure \ref{example_2} 
we plot the real and imaginary parts of the calculated phase
function outside the star versus the compactified rotated coordinate x. 

In the next sections we will show the results obtained for realistic equations
of state.
\begin{table}\footnotesize
\caption{Frequency and damping time of the trapped modes (t), the spatial
  modes and the interface 
  modes (i), in central density units, for axial l=2
  modes, $2M/R = 
  0.885$.}
\begin{tabular}{ c c c c }
  \hline \hline
  $\omega_{\Re}$ & $\omega_{\Im}$ & $\omega_{\Re}$ & $\omega_{\Im}$ \\
  \hline  
 (t)$0.21386387$ & $2.4320459 \cdot 10^{-9}$ & $1.68347876$ & $6.0243716
 \cdot 10^{-2}$  \\ 
 (t)$0.29101153$ & $7.7472523 \cdot 10^{-8}$ & $1.76704836$ & $6.1930099
 \cdot 10^{-2}$  \\ 
 (t)$0.36799864$ & $1.0725279 \cdot 10^{-6}$ & $1.85066583$ & $6.3390024
 \cdot 10^{-2}$  \\ 
 (t)$0.44463493$ & $9.5187522 \cdot 10^{-6}$ & $1.93429817$ & $6.4699961
 \cdot 10^{-2}$  \\ 
 (t)$0.52063903$ & $6.3393194 \cdot 10^{-5}$ & $2.01793447$ & $6.5897686
 \cdot 10^{-2}$  \\ 
 $0.59557900$ & $3.3806006 \cdot 10^{-4}$ & $2.10157798$ & $6.7019361
 \cdot 10^{-2}$  \\ 
 $0.66896067$ & $1.4322666 \cdot 10^{-3}$ & $2.18522398$ & $6.8080510
 \cdot 10^{-2}$  \\ 
 $0.74090953$ & $4.5080077 \cdot 10^{-3}$ & $2.26888071$ & $6.9098878
 \cdot 10^{-2}$  \\ 
 $0.81280272$ & $1.0135742 \cdot 10^{-2}$ & $2.35254774$ & $7.0076042
 \cdot 10^{-2}$  \\
 $0.88590397$ & $1.7259264 \cdot 10^{-2}$ & $2.43622599$ & $7.1018102
 \cdot 10^{-2}$  \\ 
 $0.96037307$ & $2.4480195 \cdot 10^{-2}$ & (i)$1.45115909$ & $5.4011419
 \cdot 10^{-2}$  \\ 
 $1.03608110$ & $3.1027484 \cdot 10^{-2}$ & (i)$1.73433905$ & $1.3945066
 \cdot 10^{-2}$  \\ 
 $1.11308152$ & $3.6685221 \cdot 10^{-2}$ & (i)$2.01106806$ & $2.2902687
 \cdot 10^{-2}$  \\ 
 $1.19147020$ & $4.1559313 \cdot 10^{-2}$ & (i)$2.29393679$ & $3.1975625
 \cdot 10^{-2}$  \\ 
 $1.27122330$ & $4.5838031 \cdot 10^{-2}$ & (i)$2.57982976$ & $4.1067802
 \cdot 10^{-2}$  \\ 
 $1.35219268$ & $4.9636532 \cdot 10^{-2}$ & (i)$2.86633683$ & $5.0155809
 \cdot 10^{-2}$  \\ 
 $1.43415622$ & $5.2976144 \cdot 10^{-2}$ & (i)$3.15274466$ & $5.9234204
 \cdot 10^{-2}$  \\ 
 $1.51685560$ & $5.5841641 \cdot 10^{-2}$ & (i)$3.43712011$ & $6.8307813
 \cdot 10^{-2}$  \\ 
 $1.60003382$ & $5.8244121 \cdot 10^{-2}$ \\
  \hline \hline
\end{tabular}
\label{tab:test}
\end{table}
\section{Results for realistic equations of state}
In this section we present our results for the axial quasi-normal modes of
neutron stars with realistic equations of state. We consider a wide range of equations of state in order to 
study the influence of exotic matter in the core of the star on the axial quasi-normal modes. 

Using the parametrization presented by Read et all \cite{Jocelyn2009}, which is implemented in our code, we can study the 
34 equations of
state they considered. For this paper we have used, following their notation, SLy, APR4, BGN1H1, GNH3, H1, H4, ALF2, ALF4.

After the recent measurement of the $1.97 M_\odot$ for the pulsar PSR
J164-2230, several equations of state have been proposed. These new equations
allow the presence of exotic matter in the core of the neutron star, and the
maximum mass for stable configurations is over this value. We have  
considered the following ones: two equations of state presented by
Weissenborn et al with hyperons in \cite{Weissen1},  we call them  WCS1 y WCS2;
three EOS presented by  Weissenborn et al with quark matter in
\cite{Weissen2}, we call them WSPHS1, WSPHS2,  WSPHS3; four equations of state
presented by L. Bonanno and A. Sedrakian in \cite{Sedrakian}; we call them BS1,
BS2, BS3, BS4; and one EOS presented by Bednarek et al in \cite{Bednarek}, we
call it BHZBM.
\begin{figure}
\includegraphics[angle=-90,width=0.45\textwidth]{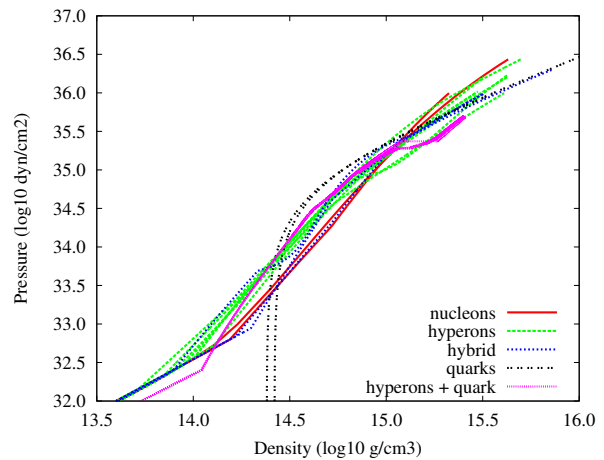} 
\caption{Pressure versus density in logarithmic scale for the 18
  equations of state considered, in the high density region.}
\label{eos_full}
\end{figure}

In Figure \ref{eos_full} we plot all the 18 equations of state we have
studied in order to give an idea of the range considered. Let us describe briefly the characteristics of the equations of state we have used: 

For plain $npe\mu$ nuclear matter, we use:
\begin{itemize}
\item SLy \cite{Haensel2001_SLy} with $npe\mu$ using a potential-method to obtain the eos.
\item APR4 \cite{akmal_1998} with $npe\mu$, obtained using a variational-method.
\end{itemize}
For mixed hyperon-nuclear matter we use:
\begin{itemize}
\item GNH3 \cite{Glendenning1985}, a relativistic mean-field theory eos
  containing hyperons.
\item H1 and H4 \cite{lackey_2006}, two variants of the GNH3 equation of state.
\item BGN1H1 \cite{Balberg}, an effective-potential equation of state including hyperons.
\item WCS1 and WCS2 \cite{Weissen1}, two equations of state with
  hyperon matter, using ``model
  $\sigma\omega\rho\phi$'', and considering ideal mixing, SU(6) quark model, and
  the symmetric-antisymmetric couplings ratio $\alpha_v=1$ and $\alpha_v=0.2$
  respectively. 
\item BHZBM \cite{Bednarek}, a non-linear relativistic mean field model
  involving baryon octet coupled to meson fields.
\end{itemize}
For Hybrid stars we use:
\begin{itemize}
\item ALF2 and ALF4 \cite{Alford2005}, two hybrid EOS with mixed APR nuclear
  matter and color-flavor-locked quark matter.
\item WSPHS3 \cite{Weissen2}, a hybrid star calculated using bag
  model, mixed with NL3 RMF hadronic EOS. The parameters employed are:
  $B_{eff}^{1/4}=140 MeV$, $a_4=0.5$, and a Gibbs phase transition.
\end{itemize}
For Hybrid stars with hyperons and quark color-superconductivity we use:
\begin{itemize}
\item BS1, BS2, BS3, and BS4, \cite{Sedrakian}, four equations of state
  calculated using a combination of phenomenological relativistic
  hyper-nuclear density functional and an effective NJL model of quantum
  chromodynamics. The parameters considered are: vector coupling $G_V/G_S=0.6$
  and quark-hadron transition density $\rho_{tr}/\rho_{0}$ equal to
  $2$, $3$, $3.5$ and $4$ respectively, where $\rho_0$ is the density of nuclear saturation.
\end{itemize}
For quark stars we use:
\begin{itemize}
\item WSPHS1 and WSPHS2 \cite{Weissen2}. The first equation of
  state is for unpaired quark matter, and we have considered the parameters
  $B_{eff}^{1/4}=123.7 MeV$,$a_4=0.53$. The second equation of state considers
  quark matter in the CFL phase (paired). The parameters considered are
  $B_{eff}^{1/4}=130.5 MeV$,$a_4=0.66$,$\Delta=50 MeV$
\end{itemize}
\begin{table*}
\footnotesize
\caption{Fits for the wI0 modes. Parameters $A$ and $B$ correspond to the
  linear 
  empirical relation for the frequency (\ref{empiricalfrec}). Parameters $a$,$b$
  and $c$ correspond to the quadratic empirical relation for the damping time
  (\ref{empiricaltau}).}
\begin{tabular*}{\linewidth}{ c c c c c c c c c c}
    \hline \hline
wI0 & SLy & APR4 & GNH3 & BGN1H1 & H1 & H4 & WCS2 & WCS1 & BHZBM
 \\
  \hline 
$A$         &$-148.7\pm4.5$&$-164.2\pm5.8$&$-122.1\pm1.4$&$-95.5\pm5.4$&$-89.9\pm5.1$&$-120.6\pm1.8$&$-139.7\pm3.6$&$-107.1\pm2.2$&$-132.8\pm3.6$\\
$B$         &$119.8\pm1.0$ &$122.4\pm1.4$&$116.67\pm0.25$&$111.9\pm1.0$&$113.06\pm0.81$&$115.50\pm0.35$&$117.83\pm0.75$&$112.81\pm0.43$&$117.35\pm0.78$\\
$\chi^{2}$&0.760&0.039&1.56&0.689&0.260&0.0812&0.430&0.096&0.388\\
  \hline  
$a$          &$-1221\pm22$&$-1112\pm19$&$-1444\pm23$&$-1690\pm49$&$-1539\pm42$&$-1385\pm47$&$-1244\pm23$&$-1450\pm64$&$-1401\pm41$\\
$b$          &$365.1\pm9.3$&$331.1\pm9.0$&$425.4\pm8.3$&$506\pm18$&$448\pm13$&$414\pm17$&$371.4\pm9.6$&$437\pm24$&$418\pm17$\\
$c$          &$21.63\pm0.89$&$24.16\pm0.94$ &$17.87\pm0.68$&$11.2\pm1.6$&$16.6\pm1.0$&$18.2\pm1.5$&$21.4\pm0.9$&$16.2\pm2.1$&$17.9\pm1.6$\\
$\chi^{2}$ &$0.050$&$0.034$&$0.027$&$0.084$&$0.016$&$0.091$&$0.044$&$0.083$&$0.097$\\
  \hline \hline
\end{tabular*}
\label{tab:wI0fit_hyp}
\end{table*}
\begin{table*}
\footnotesize
\caption{Fits for the wI1 modes. Parameters $A$ and $B$ correspond to the
  linear 
  empirical relation for the frequency (\ref{empiricalfrec}). Parameters $a$,$b$
  and $c$ correspond to the quadratic empirical relation for the damping time (\ref{empiricaltau}).}
\begin{tabular*}{\linewidth}{ c c c c c c c c c c}
    \hline \hline
 wI1  & SLy & APR4 & GNH3 & BGN1H1 & H1 & H4 & WCS2 & WCS1 & BHZBM
 \\
  \hline  
$A$         &$-689.5\pm5.2$&$-667.9\pm3.9$&$-736.5\pm6.8$&$-779.4\pm6.8$&$-797\pm12$&$-703.7\pm5.5$&$-684.2\pm6.8$&$-723.5\pm7.0$&$-727.7\pm6.2$\\
$B$         &$312.5\pm1.1$&$309.35\pm0.97$&$317.2\pm1.4$&$323.3\pm1.4$&$324.4\pm0.7$&$316.2\pm1.1$&$310.5\pm1.5$&$315.3\pm1.4$&$317.5\pm1.3$\\
$\chi^{2}$&  0.997           &  0.670           &  0.372          &  1.986& 0.18& 0.731& 1.564& 0.955& 1.144\\
  \hline  
$a$          &$-1843\pm 55$&$-1792\pm65$&$-2062\pm 25$&$-1771\pm 123$&$-2274\pm20$&$-2020\pm71$&$-1863\pm78$&$-2635\pm150$&$-1854\pm79$\\
$b$          &$650\pm23$&$646\pm30$&$690.9\pm 8.9$&$582\pm45$&$725.8\pm6.6$&$687\pm26$&$656\pm32$&$879\pm55$&$627\pm32$\\
$c$         &$14.3\pm2.2$&$13.7\pm 3.2$&$13.10\pm 0.73$&$21.1\pm 3.9$&$12.02\pm 0.51$&$12.7\pm 2.3$&$14\pm 29$&$3.1\pm 4.8$&$17.4\pm 3.0$\\
$\chi^{2}$ &$0.311$&$0.510$&$0.380$&$0.393$&$0.00374$&$0.209$&$0.401$&$0.520$&$0.361$\\
  \hline \hline
\end{tabular*}
\label{tab:wI1fit_hyp}
\end{table*}
\subsection{Nuclear and Hyperon matter}
We will start studying the impact of hyperon matter on the quasi-normal mode
spectrum, comparing with configurations with plain nuclear matter EOS.

\begin{figure}
\includegraphics[angle=-90,width=0.45\textwidth]{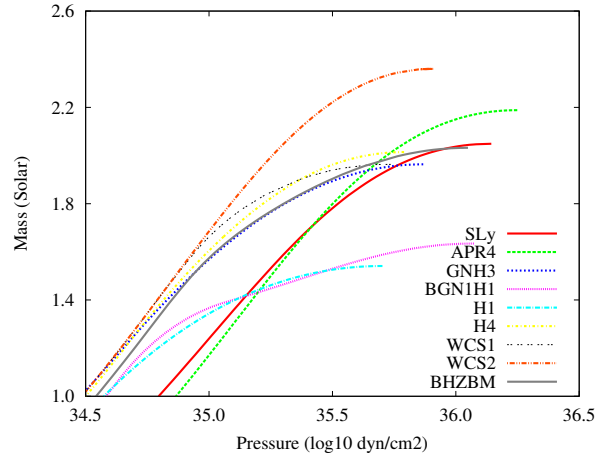} 
\caption{Mass vs central pressure for the equations of state considered in
  Section V.A. We plot configurations up to the maximum stable mass. Note
  that all the EOS have a maximum mass nearby or beyond 
  $2M_{\odot}$, except H1 and BGN1H1, which are included for comparison.}
\label{eos_M_pc_hyperons}
\end{figure}
\begin{figure}
\includegraphics[angle=-90,width=0.45\textwidth]{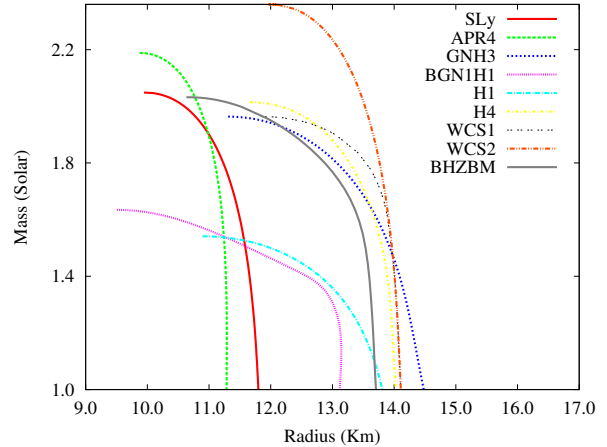} 
\caption{Mass vs Radius for the equations of state  considered in
  Section V.A. Typically, stars with hyperon matter have bigger radius than nuclear
  matter stars of the same mass.}
\label{M_R_hyperons}
\end{figure}
In Figures \ref{eos_M_pc_hyperons} and \ref{M_R_hyperons} we plot the
mass-central pressure and
mass-radius relation, respectively.
For
the study of the quasi-normal modes we have considered only stable
configurations below 
the maximum mass of each equation of state. It can be seen that the
introduction of a hyperon core 
changes the mass radius relation and we will see that this has an imprint
on the quasi-normal mode spectrum of these configurations. Note that all the
EOS considered are near or beyond the $1.97 M_\odot$ limit,
except the BGN1H1 and H1 equations of state, which are included for
comparison. In the following we will consider only stable configurations between $1
M_\odot$ and the maximum mass of the particular equation of state. We will
study the fundamental and first excited wI modes, as well as the fundamental
wII mode.

 \begin{figure}
 \includegraphics[angle=-90,width=0.45\textwidth]{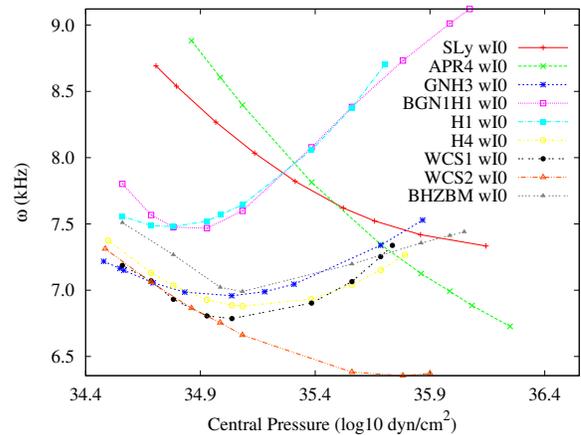} 
 \caption{Frequency of the fundamental wI mode vs central pressure in
   logarithmic scale, for the EOS considered in
  Section V.A. While there is no significant difference between the
   SLy and the APR4, the hyperon core
   changes the relation of the frequency with the central pressure for high densities ($p_c$ above 
   $\sim 10^{35} dyn/cm^2$)}  
 \label{eos_wI0_real_pc}
 \end{figure}
  \begin{figure}
  \includegraphics[angle=-90,width=0.45\textwidth]{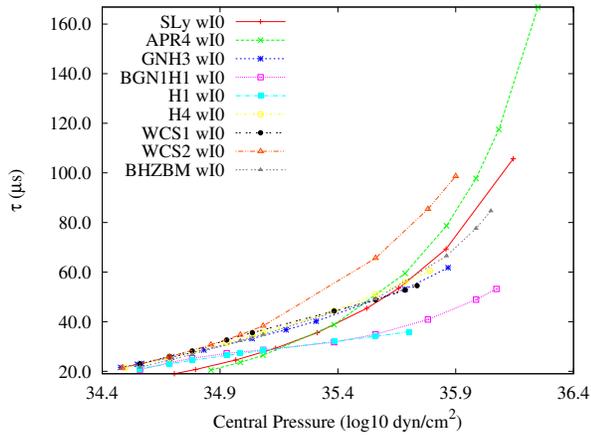}
  \caption{Damping time of the fundamental wI mode vs central pressure in
   logarithmic scale, for the EOS considered in
  Section V.A. The behavior is quite similar for all the equations of
   state considered. At high density, the lowest values of the damping time are reached by the
   H1 and BGN1H1 EOS.} 
  \label{eos_wI0_imaginary_pc}
  \end{figure}

\textbf{Fundamental wI mode}: In Figures \ref{eos_wI0_real_pc} and \ref{eos_wI0_imaginary_pc}, we plot the
frequency and damping time of the fundamental wI
mode as a function of the central pressure of the star in logarithmic scale, for the different
equations of state. There is no significant difference between the two
pure nuclear 
matter equations of state considered, except for the bigger maximum mass of
the APR4 EOS. On the other hand, we observe that the
presence of hyperon condensates has a clear impact on the frequencies of the
fundamental 
w-modes. Note that the hyperon core is changing the frequency
relation for high mass configurations, increasing with the central pressure
above $\sim 10^{35} dyn/cm^2$, where the hyperon phases are more important due
to the high density of the star core. Essentially, this effect is related to
the softening of the EOS by the introduction of hyperon
matter at the inner regions of the star. 

This result is in agreement with a recent work by Chatterjee and Bandyopadhyay
\cite{Chatterjee2009}. 

 \begin{figure}
 \includegraphics[angle=-90,width=0.45\textwidth]{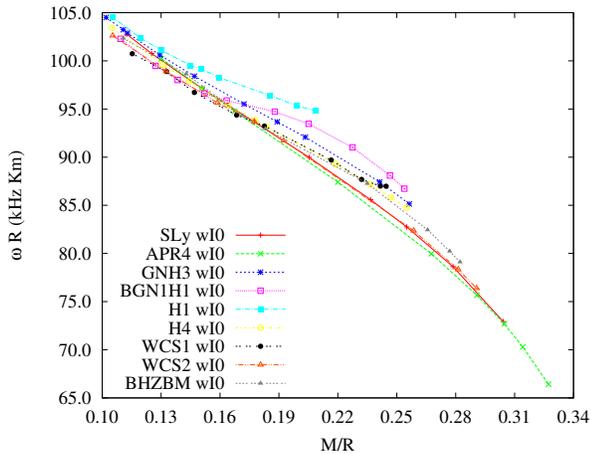}  
 \caption{Scaled frequency of the fundamental wI mode vs M/R, for the EOS considered in
  Section V.A. Except for H1
   and BGN1H1 EOS, which present a quite different behavior, all the scaled
   frequencies can be fitted to a linear relation 
   with the compactness (\ref{empiricalfrec}). The fits can be found in
   Table \ref{tab:wI0fit_hyp}.} 
 \label{eos_wI0_real_M_R_scaled_hyp}
 \end{figure}
  \begin{figure}
  \includegraphics[angle=-90,width=0.45\textwidth]{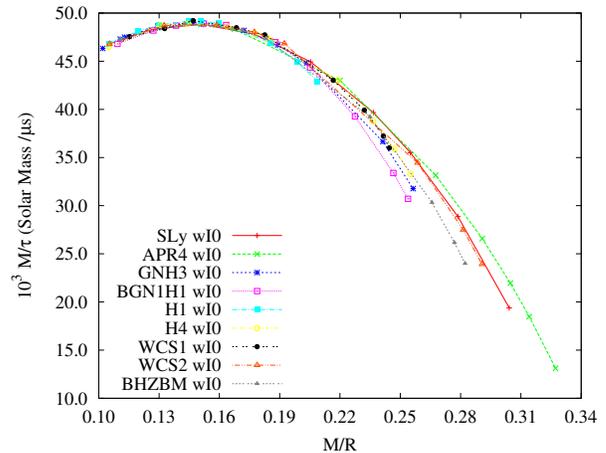}
  \caption{Scaled damping time of the fundamental wI mode vs M/R, for the EOS considered in
  Section V.A. The scaled damping times can be fitted to a quadratic
    relation with the 
    compactness (\ref{empiricaltau}). The fits can be found in Table
    \ref{tab:wI0fit_hyp}.} 
  \label{eos_wI0_imaginary_M_R_scaled_hyp}
  \end{figure}

In order to use future observations of gravitational waves to estimate  the
mass and the radius of the neutron star, as well as  
 to discriminate between different families of equations of state, we obtain
 empirical relations between the frequency and damping time of 
quasi-normal modes and the compactness of the star, following
\cite{Andersson_obs_1996}, \cite{Kokkotas2001},\cite{Kokkotas_asteroseismology_1998}
and \cite{Benhar1999}. In Figure 
\ref{eos_wI0_real_M_R_scaled_hyp} we present  the
frequency of the fundamental mode scaled to the radius of each
configuration. It is interesting to note that even with this scaling, 
the
softest equations of state that include hyperon matter, H1 and BGN1H1, present a quite
different behavior than the rest of EOS considered. Nevertheless,
as the detection of the recent $2 M_{\odot}$ pulsar suggest, these two
particular EOS cannot be realized in nature, so the behavior of the scaled
frequency is quite similar for hyperon matter or plain nuclear matter stars. 

A linear fit can be made to each equation of state. We fit to the following
phenomenological relation:
\begin{equation}
\omega(Khz) = \frac{1}{R(Km)}\left(A\frac{M}{R} + B\right).
\label{empiricalfrec}
\end{equation}
In Table \ref{tab:wI0fit_hyp} we present the fit parameters $A$,$B$ for each
one of the equations of state studied.
For the Sly and for the APR4 equations very similar results are obtained.
The empirical parameters for SLy and APR4 are compatible
with the empirical relation  obtained by Benhar et al \cite{Benhar1999} for
six equations of state. 

In general, for hyperon matter EOS, we still obtain a linear relation, but
with slightly 
different parameters, as can be seen in Table \ref{tab:wI0fit_hyp}. As
commented before, quite
remarkable is the case of the H1 and BGN1H1, where the linear relation is
almost lost.  

Our results show that the introduction of
hyperon matter changes very little the 
dependence of the frequency with the compactness. 
Even for $M/R$ bigger than
0.18, the difference between equations of state with and without hyperon
matter is quite small. As these empirical relations depend minimally on the
equation of state, they could be used, applying the technique from
\cite{Kokkotas_asteroseismology_1998} to measure the radius of the neutron
star and constrain the equation of state.   

In Figure
\ref{eos_wI0_imaginary_M_R_scaled_hyp} we present the
damping time of the fundamental mode scaled to the mass of each
configuration. 
In this case, the results can be fitted to an empirical quadratic relation on $M/R$, as follows:
\begin{equation}
\frac{10^3}{\tau(\mu s)} = \frac{1}{M(M_{\odot})}\left[a\left(\frac{M}{R}\right)^{2} + b\frac{M}{R} + c\right].
\label{empiricaltau}
\end{equation}    

In Table \ref{tab:wI0fit_hyp} we present the fit parameters a,b and c. Note
that for  SLy and APR4 the fits are quite similar, and in accordance with the
results obtained in \cite{Benhar1999}. All of the equations of state are well
fitted to a quadratic relation with the compactness. For compactness beyond
0.18 the presence of hyperon matter slightly changes the 
dependence of the scaled damping time with $M/R$, specially for the BGN1H1
EOS.

 \begin{figure}
 \includegraphics[angle=-90,width=0.45\textwidth]{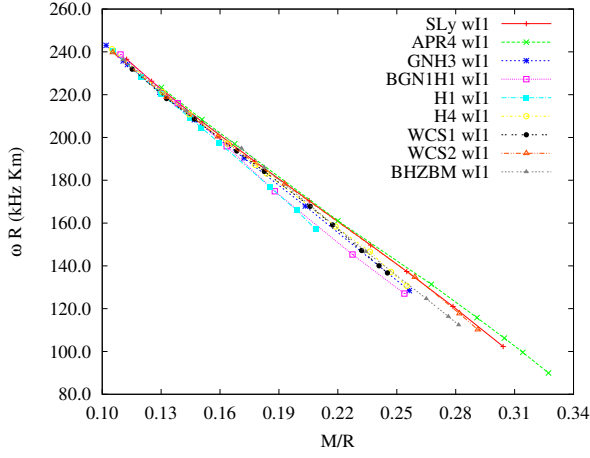} 
 \caption{Scaled frequency of the first excited wI mode vs M/R, for the EOS considered in
  Section V.A. In this case
   the linear relation (\ref{empiricalfrec}) is found for all these EOS. If we exclude the H1 and
   BN1H1 EOS, the presence of hyperons is not affecting strongly the scaled
   frequency. The
   fits can be found in 
   Table \ref{tab:wI1fit_hyp}.}
 \label{eos_wI1_real_M_R_scaled_hyp}
 \end{figure}
  \begin{figure}
  \includegraphics[angle=-90,width=0.45\textwidth]{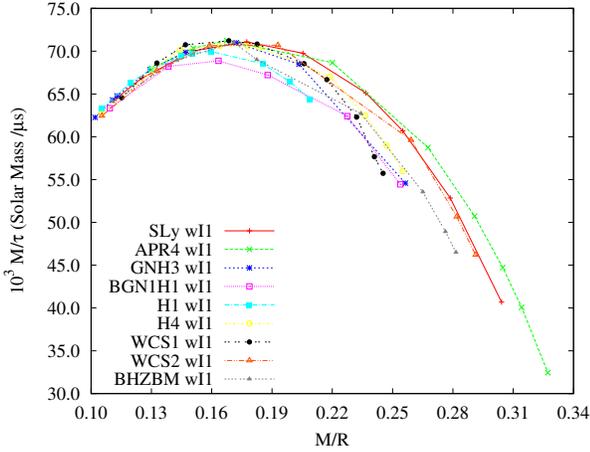}
  \caption{Scaled damping time of the first excited wI mode vs M/R, for the EOS considered in
  Section V.A. The
    quadratic relation (\ref{empiricaltau}) is valid for every EOS considered. The empirical parameters are
    more sensitive to the hyperon core, specially at high densities. The
   fits can be found in 
   Table \ref{tab:wI1fit_hyp}.}
  \label{eos_wI1_imaginary_M_R_scaled_hyp}
  \end{figure}

\textbf{First excited wI mode}: In Figures
\ref{eos_wI1_real_M_R_scaled_hyp} and \ref{eos_wI1_imaginary_M_R_scaled_hyp} we plot
the scaled frequency and damping time for the first excited wI modes. We make
fits to the empirical 
relations (\ref{empiricalfrec}) and (\ref{empiricaltau}). In Table
\ref{tab:wI1fit_hyp} we present the corresponding empirical parameters.



 \begin{figure}
 \includegraphics[angle=-90,width=0.45\textwidth]{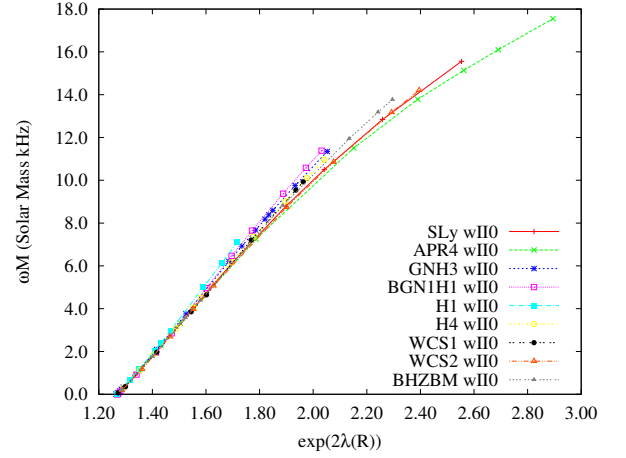} 
 \caption{Scaled frequency of the fundamental wII mode vs
   $e^{2\lambda(R)}$, for the EOS considered in
  Section V.A. The frequency vanishes for compactness between
$M/R=0.1044$ and $M/R=0.1066$. For low compactness, there are no or almost no hyperons in the core of the star, so at
this point all the configurations are basically plain nuclear matter stars.} 
 \label{eos_wII0_real_M_R_scaled_hyp}
 \end{figure}
  \begin{figure}
  \includegraphics[angle=-90,width=0.45\textwidth]{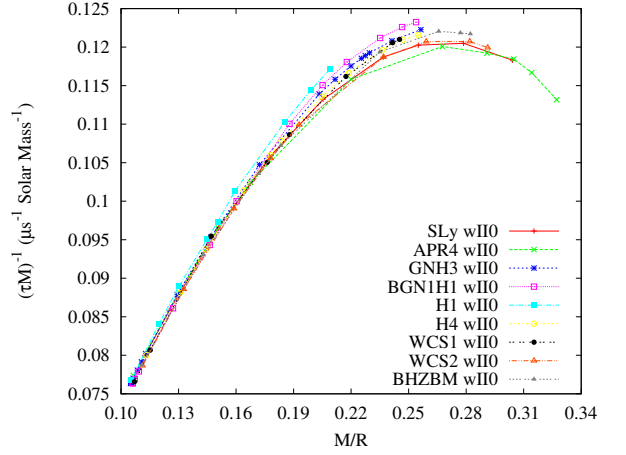}
  \caption{Scaled damping time of the fundamental wII mode vs M/R, for the EOS considered in
  Section V.A. The damping time
    does not vanish in the limit compactness as the frequency does. }
  \label{eos_wII0_imaginary_M_R_hyp}
  \end{figure}

\textbf{Fundamental wII mode}: In Figure
\ref{eos_wII0_real_M_R_scaled_hyp} we plot the frequency of the fundamental wII
mode scaled with the mass versus the metric function $e^{2\lambda(R)}$. It can be
seen that for all the equations of state studied, there is a minimal
compactness, below which the wII mode does not exist. The scaled frequency is
quite linear with $e^{2\lambda(R)}$ nearby this minimal compactness. Our
numerical scheme allows us to obtain explicitly these limit configurations with
vanishing  
 real part of the mode with a precision of the order of $10^{-5}$. The limit
 configuration has compactness between
$M/R=0.1044$ and $M/R=0.1066$, so it is quite independent of the particular
equation of state, and in consequence independent of the presence of
hyperon matter.  For example, the GNH3 equation of state the limit
configuration has 
compactness $M/R=0.1051$, and for the BGN1H1, $M/R=0.1061$, and the SLy has
compactness $M/R=0.1056$ and the APR4 compactness $M/R=0.1046$. The reason
why the limit configuration is independent of the EOS is because, for these
values of the compactness, the mass of the stars 
considered are below $1 M_{\odot}$. In these low mass configurations the
presence of hyperons at the core is small or null, depending of the equation
of state, and so we are comparing plain nuclear matter stars with slightly
different EOS at the core. 


These limit configurations were conjectured by extrapolation in Wen et al in
\cite{Wen2009}. 
Our results are in agreement with their extrapolation.

In Figure \ref{eos_wII0_imaginary_M_R_hyp} we show the damping time of
the wII mode scaled with the mass. 
Although for low compactness the damping time is quite independent of the
equation of state and linear in $M/R$ for the reasons commented before, for
compactness above 0.18 it is very 
sensitive to the presence of 
hyperon matter, changing completely the linear behavior. 
The damping time does not vanish in the limit compactness as the
frequency does. By the contrary, it tends to a limit $\tau$ 
. 
\begin{figure}
\includegraphics[angle=-90,width=0.45\textwidth]{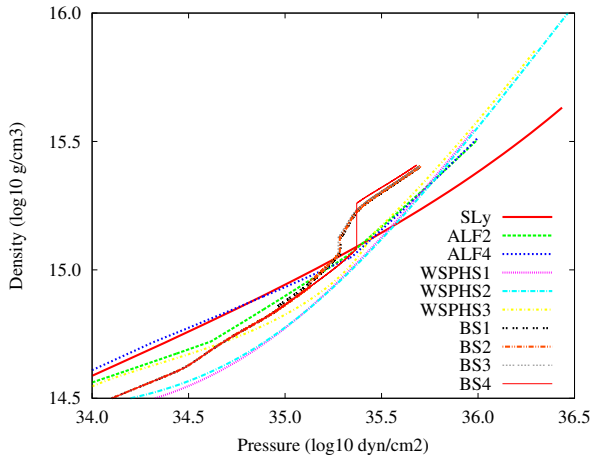} 
\caption{Density versus pressure in logarithmic scale in the high density region, for the EOS considered in
  Section V.B. Note that hybrid stars
  present important phase transitions at high pressures.}
\label{eos_zoom_quark}
\end{figure}
\begin{figure}
\includegraphics[angle=-90,width=0.45\textwidth]{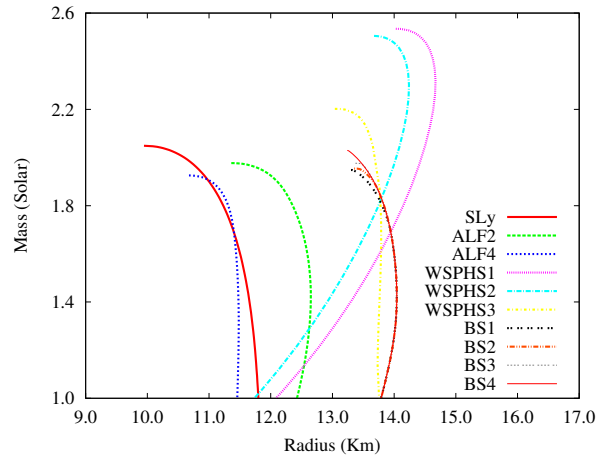} 
\caption{Mass vs Radius relation for the EOS considered in
  Section V.B. All
  these EOS have maximum mass nearby or beyond $2 M_{\odot}$ and radius beyond $10 Km$.}
\label{eos_MR_quark}
\end{figure}
\begin{figure}
\includegraphics[angle=-90,width=0.45\textwidth]{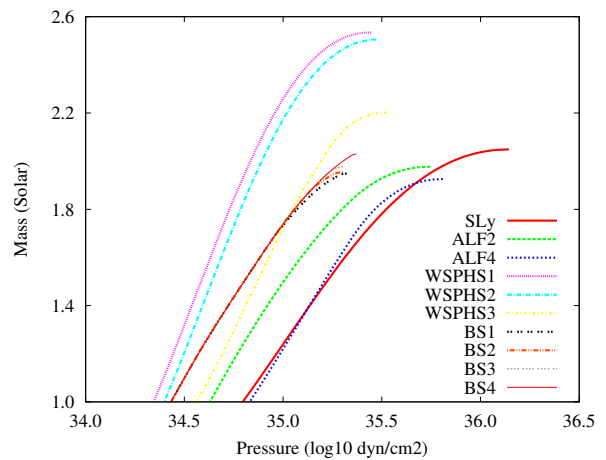} 
\caption{Mass vs central pressure relation, for the EOS considered in
  Section V.B. In this study we consider only stable configurations
  between $1 M_{\odot}$ and the maximum mass of the particular EOS.}.
\label{eos_p_quark}
\end{figure}   
\begin{table*}
\footnotesize
\caption{Fits for the wI0 modes. Parameters $A$ and $B$ correspond to the
  linear 
  empirical relation for the frequency (\ref{empiricalfrec}). Parameters $a$,$b$
  and $c$ correspond to the quadratic empirical relation for the damping time (\ref{empiricaltau}).}
\begin{tabular*}{\linewidth}{ c c c c c c c c c c}
    \hline \hline
 wI0  & ALF2 & ALF4 & WSPHS1 & WSPHS2 & WSPHS3 & BS1 & BS2 & BS3 & BS4
 \\
  \hline  
$A$         &$-119.0\pm3.0$&$-141.4\pm1.2$&$-56.5\pm3.7$&$-57.95\pm3.1$&$-145.3\pm3.4$&$-102.1\pm1.8$&$-104.3\pm1.7$&$-104.9\pm1.4$&$-104.7\pm1.1$\\
$B$         &$113.51\pm0.59$ &$118.40\pm0.25$&$94.31\pm0.75$&$93.96\pm0.65$&$119.98\pm0.66$&$110.14\pm0.32$&$110.42\pm0.31$&$110.50\pm0.25$&$110.46\pm0.20$\\
$\chi^{2}$&0.151&0.039&0.182&0.182&0.178&0.0359&0.0343&0.0227&0.0174\\
  \hline  
$a$          &$-1290\pm42$&$-1158\pm26$&$-1319\pm75$&$-295\pm35$&$-1147\pm24$&$-1275\pm46$&$-1263\pm46$&$-1266\pm43$&$-1283\pm34$\\
$b$          &$395\pm16$&$345.3\pm9.8$&$455\pm30$&$445\pm14$&$340.4\pm8.7$&$394\pm15$&$390\pm15$&$391\pm14$&$397\pm11$\\
$c$          &$18.6\pm1.4$&$23.06\pm0.88$ &$9.3\pm2.3$&$10.4\pm1.4$&$23.62\pm0.76$&$18.4\pm1.1$&$18.6\pm1.1$&$18.5\pm1.1$&$18.14\pm0.90$\\
$\chi^{2}$ &$0.043$&$0.028$&$0.134$&$0.044$&$0.017$&$0.025$&$0.025$&$0.023$&$0.021$\\
  \hline \hline
\end{tabular*}
\label{tab:wI0fit_quark}
\end{table*}
\begin{table*}
\footnotesize
\caption{Fits for the wI1 modes. Parameters $A$ and $B$ correspond to the
  linear 
  empirical relation for the frequency (\ref{empiricalfrec}). Parameters $a$,$b$
  and $c$ correspond to the quadratic empirical relation for the damping time (\ref{empiricaltau}).}
\begin{tabular*}{\linewidth}{ c c c c c c c c c c}
    \hline \hline
 wI1 & ALF2 & ALF4 & WSPHS1 & WSPHS2 & WSPHS3 & BS1 & BS2 & BS3 & BS4
 \\
  \hline  
A         &$-715.8\pm4.0$&$-688.7\pm6.8$&$-560\pm19$&$-562\pm18$&$-672.6\pm6.7$&$-685\pm19$&$-694\pm10$&$-689\pm11$&$-681\pm11$\\
B         &$316.98\pm0.78$ &$313.5\pm1.4$&$283.0\pm4.0$&$283.3\pm3.8$&$310.0\pm1.3$&$308.9\pm3.5$&$310.4\pm2.0$&$309.7\pm2.0$&$308.3\pm2.0$\\
$\chi^{2}$&  0.258           &  1.236           &  5.073          &  6.203& 0.691& 1.821& 0.6037& 0.662& 0.921\\
  \hline  
 
a          &$-1828\pm 72$&$-1857\pm110$&$-2026\pm 103$&$-2017\pm 117$&$-1870\pm79$&$-1536\pm142$&$-1874\pm155$&$-1897\pm205$&$-1458\pm85$\\
b          &$637\pm27$&$654\pm42$&$780\pm 40$&$785\pm47$&$639\pm29$&$549\pm50$&$661\pm55$&$670\pm73$&$524\pm32$\\
c          &$14.7\pm2.5$&$14.2\pm 3.8$&$-5.1\pm 3.9$&$6.3\pm 4.5$&$17.9\pm 2.5$&$20.9\pm 4.3$&$11.8\pm 4.7$&$11.0\pm 6.3$&$22.8\pm 2.9$\\
$\chi^{2}$ &$0.1242$&$0.510$&$0.252$&$0.486$&$0.187$&$0.0502$&$0.0584$&$0.1167$&$0.0457$\\
  \hline \hline
\end{tabular*}
\label{tab:wI1fit_quark}
\end{table*}
\subsection{Quark matter}
In this subsection we will present our results for the axial quasi-normal
modes of compact stars with quark  
matter at the core. We consider the following equations of state: ALF2, ALF4, WSPHS1-3,
BS1-4, already described previously. In the following Figures we will always
include the results 
for the SLy EOS for comparison. 

In Figure \ref{eos_zoom_quark} we present the equations of state
at the high density 
region.
In Figure \ref{eos_MR_quark}  we present the mass-radius relation for these
equations of state and in Figure \ref{eos_p_quark} the mass-central pressure relation.
Only stable 
configurations below the maximum mass of each equation of state have been considered.  
Note that all the
equations of state considered are near or above  the $1.97 M_\odot$ limit. We
will repeat the analysis for these new equations, studying the fundamental and first
excited wI modes, and finally the fundamental wII mode.

 \begin{figure}
 \includegraphics[angle=-90,width=0.45\textwidth]{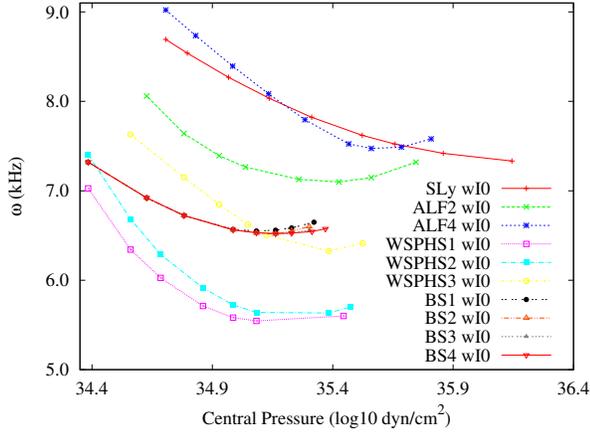} 
 \caption{Frequency of the fundamental wI mode vs central pressure in
   logarithmic scale. Hybrid stars tend to have lower frequencies than plain nuclear matter stars or hyperon matter. Pure quak stars has the smallest frequencies, below
6 kHz.}
 \label{eos_wI0_real_pc_quark}
 \end{figure}
\begin{figure}
  \includegraphics[angle=-90,width=0.45\textwidth]{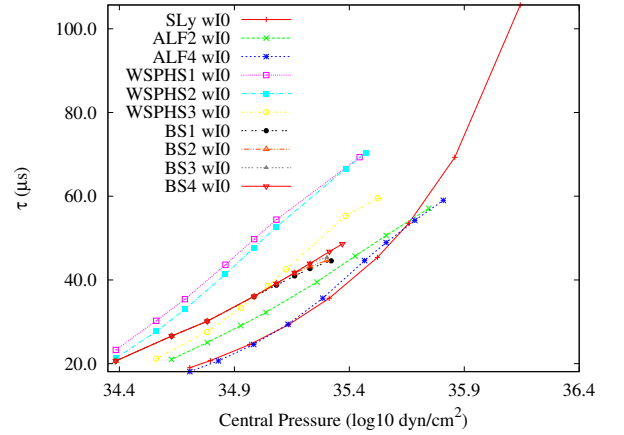}
  \caption{Damping time of the fundamental wI mode vs central pressure in
   logarithmic scale, for the EOS considered in
  Section V.B. Hybrid stars are found in between the pure quark matter
   configurations and the plain nuclear matter ones.}
  \label{eos_wI0_imaginary_pc_quark}
  \end{figure}
 \begin{figure}
 \includegraphics[angle=-90,width=0.45\textwidth]{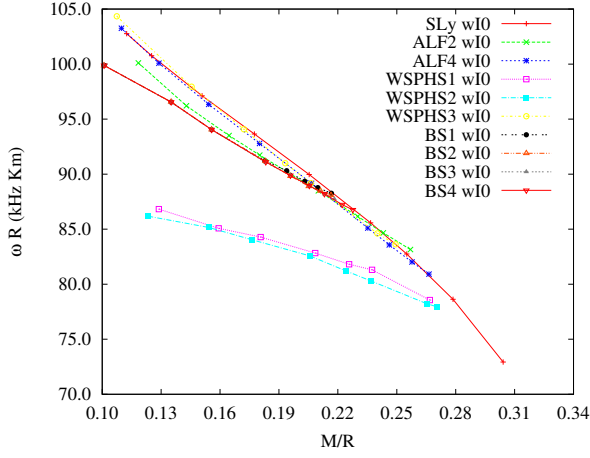}  
 \caption{Scaled frequency of the fundamental wI mode vs M/R, for the EOS considered in
  Section V.B. All the scaled
   frequencies can be fitted to a linear relation 
   with the compactness (\ref{empiricalfrec}), even the ones for WSPHS1-2. The
   empirical parameters of hybrid stars are quite different from those of pure
   quark stars. The fits can be found in Table \ref{tab:wI0fit_quark}.} 
 \label{eos_wI0_real_M_R_scaled_quark}
 \end{figure}
\begin{figure}
  \includegraphics[angle=-90,width=0.45\textwidth]{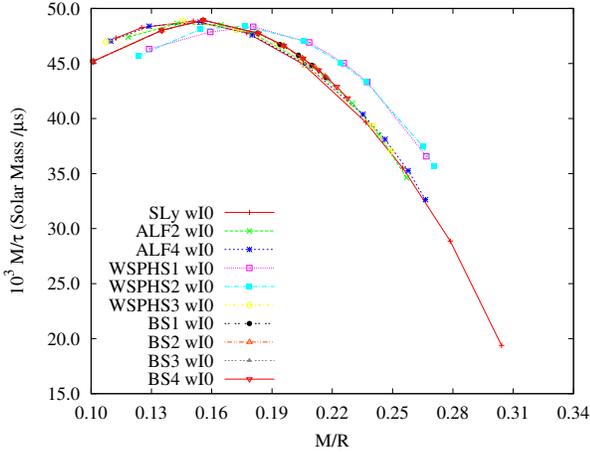}
  \caption{Scaled damping time of the fundamental wI mode vs M/R, for the EOS considered in
  Section V.B. The scaled
   damping times can be fitted to a quadratic relation 
   with the compactness (\ref{empiricaltau}). Again, WSPHS1-2 present a
   different behavior from the hybrid stars. The
   fits can be found in 
   Table \ref{tab:wI0fit_quark}.}
  \label{eos_wI0_imaginary_M_R_scaled_quark}
  \end{figure}
\textbf{Fundamental wI mode}: In Figure \ref{eos_wI0_real_pc_quark} we plot
the frequency of the fundamental wI 
mode as a function of the central pressure of the star in logarithmic scale.
In Figure \ref{eos_wI0_imaginary_pc_quark}  
 we make a similar plot for the damping time of these
configurations. Note the similitude between the SLy modes and the ALF4
modes for both the frequency and damping time. These two equations of state
can not be differentiated by the 
observation of the fundamental w mode. This is not the case if the EOS for hybrid
matter is softer, as it happens for the other EOS considered, which present
smaller frequencies.  

In Figures \ref{eos_wI0_real_pc_quark} and \ref{eos_wI0_imaginary_pc_quark} we
also plot the fundamental mode for the two pure quark matter EOS
considered (WSPHS1 and WSPHS2). The configurations near the maximum mass
limits of these EOS have much lower frequencies than the rest of the EOS
studied, around $5.5$ kHz.

In these Figures we observe that the frequency and damping times have a clear
dependence on the EOS when represented against the central pressure. All the EOS
with quark matter at the core present a characteristic behavior, with
frequencies below those of the plain nuclear matter configurations, except the
ALF4 EOS, which mimics the SLy EOS almost entirely. 

In Figure 
\ref{eos_wI0_real_M_R_scaled_quark} we present  the
frequency of the fundamental mode scaled to the radius of each
configuration. A linear fit can be made to each equation of state. We fit to the 
phenomenological relation presented in the previous subsection, equation
(\ref{empiricalfrec}). 
In Table \ref{tab:wI0fit_quark} we present the fit parameters $A$, $B$ for each
one of the equations of state studied. 

In Figure \ref{eos_wI0_imaginary_M_R_scaled_quark} we present the
damping time of the fundamental mode scaled to the mass of each
configuration. 
In this case, the results can be fitted to the empirical quadratic  relation on $M/R$, as presented in equation (\ref{empiricaltau}).
In Table \ref{tab:wI0fit_quark} we present the fit parameters a,b and c.

When we
represent the scaled frequency and damping time versus $M/R$, all the wI0
modes have a similar behavior and the coincidence is greater for higher
compactness. The only exceptions are WSPHS1-2 EOS (pure quark 
stars). Their behavior is clearly differentiated from the rest because of the
different layer structure found at the exterior of the star.
 \begin{figure}
 \includegraphics[angle=-90,width=0.45\textwidth]{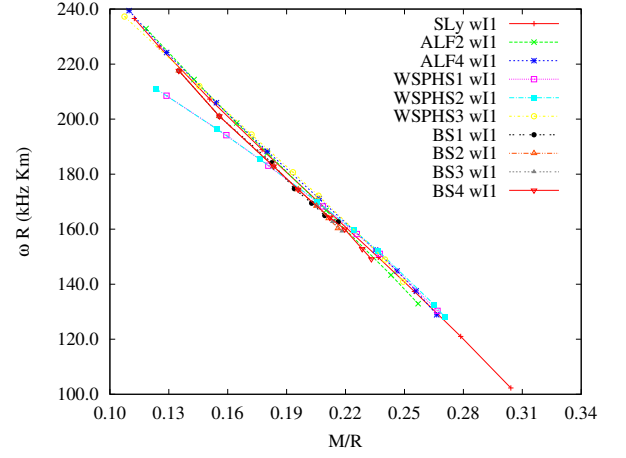} 
 \caption{Scaled frequency of the first excited wI mode vs M/R, for the EOS considered in
  Section V.B. In this case,
   all these EOS fit perfectly the linear relation 
   with the compactness (\ref{empiricalfrec}). The
   fits can be found in 
   Table \ref{tab:wI1fit_quark}.}
 \label{eos_wI1_real_M_R_scaled_quark}
 \end{figure}
  \begin{figure}
  \includegraphics[angle=-90,width=0.45\textwidth]{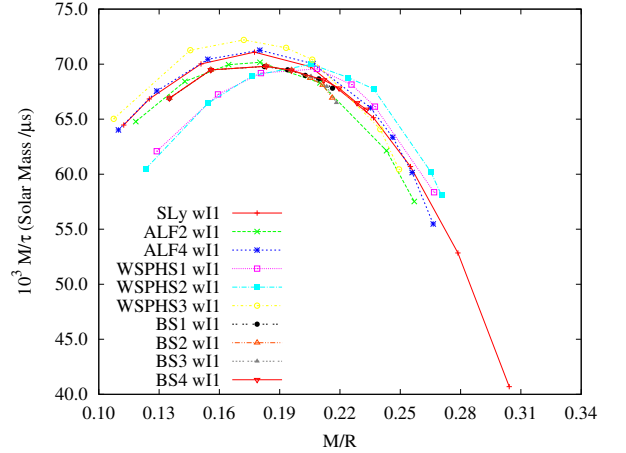}
  \caption{Scaled damping time of the first excited wI mode vs M/R, for the EOS considered in
  Section V.B. A
    quadratic relation with the compactness (\ref{empiricaltau}) can be found
    for all these EOS. For low compactness WSPHS1-2 configurations the scaled
    damping time can be a $10\%$ smaller than for hybrid stars. The
   fits can be found in 
   Table \ref{tab:wI1fit_quark}.}  
  \label{eos_wI1_imaginary_M_R_scaled_quark}
  \end{figure}

\textbf{First Excited wI mode}: We present a similar study of the first
excited wI mode. In Figures
\ref{eos_wI1_real_M_R_scaled_quark} and
\ref{eos_wI1_imaginary_M_R_scaled_quark} we plot 
the scaled frequency and damping time for the first excited wI-mode. The empirical
relations  (\ref{empiricalfrec}, \ref{empiricaltau}), with different
parameters can be found in Table \ref{tab:wI1fit_quark}. Note that the scaled
frequency and damping time is quite similar for every EOS considered, and the
only difference is found for the WSPHS1-2 EOS at low densities, for the same
reasons than for the fundamental modes.

\begin{figure}
 \includegraphics[angle=-90,width=0.45\textwidth]{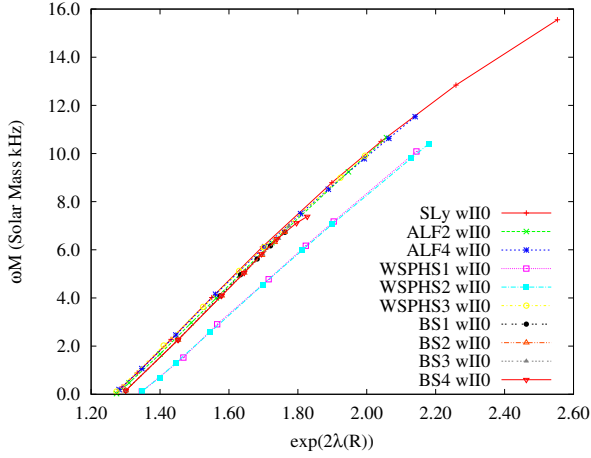} 
 \caption{Scaled frequency of the fundamental wII mode vs
   $e^{2\lambda(R)}$, for the EOS considered in
  Section V.B. The frequency vanishes for a compactness that depends on
   the lower densities of the equation of state considered, around $M/R=0.105$
   and $M/R=0.110$ for hybrid stars. The most different are the pure quark
   stars WSPHS1-2, which have the limit at $M/R=0.126$.}
 \label{eos_wII0_real_M_R_scaled_quark}
 \end{figure}

\textbf{Fundamental wII mode}: Finally, in Figure
\ref{eos_wII0_real_M_R_scaled_quark} we plot the frequency of the fundamental
wII mode scaled with the mass versus the metric function $e^{2\lambda(R)}$. As in
the 
previous subsection, it can be
seen that there is, for all the equations of state studied, a minimal
compactness below which the wII mode vanishes. 

  \begin{figure}
  \includegraphics[angle=-90,width=0.45\textwidth]{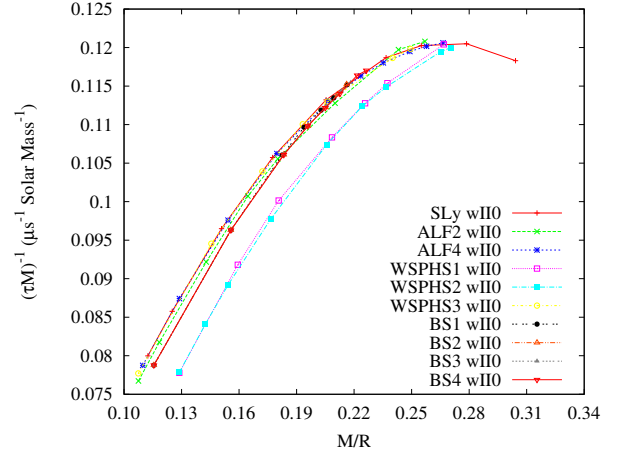}
  \caption{Damping time of the fundamental wII mode vs M/R, for the EOS considered in
  Section V.B. The damping time
    does not vanish in the limit compactness as the frequency does.}
  \label{eos_wII0_imaginary_M_R_quark}
  \end{figure}

 In Figure
\ref{eos_wII0_imaginary_M_R_quark} we show the damping time of the wII mode
scaled with the mass, where it can be seen that for these limit configurations
it tends to a limit $\tau$ dependent of the
equation of state. Note that for the hybrid stars considered, the limit
configuration is different depending on the matter composition of the outer
layers. For hybrid stars ALF2, ALF4 and WSPHS3, the limit configurations 
have compactness between 
$M/R=0.1046$ and $M/R=0.1060$, similar to hyperon stars. For BS1-4,
the limit configurations
have compactness 
$M/R=0.112$, and for quark stars WSPHS1 and WSPHS2, the limit configurations are
quite different and have compactness $M/R=0.126$. These differences are due to
the behavior of the different equations of state of each configuration at the
outer layers of the star. In fact, note that for quark stars WSPHS1 and WSPHS2
the pressure drops to zero with constant density around $2\cdot
10^{14}g/cm^{3}$. 
 \begin{figure}
 \includegraphics[angle=-90,width=0.45\textwidth]{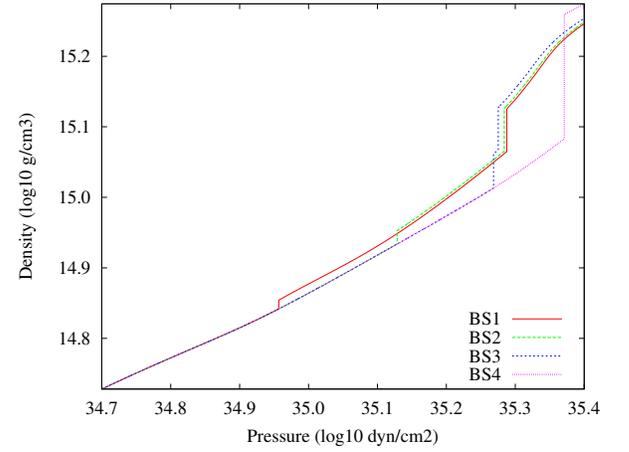} 
 \caption{Density versus pressure in logarithmic scale for the BS EOS for high densities. Note that at high pressures different phase
 transitions are taken into account, affecting the stiffness of
 the EOS.} 
 \label{eos_BS}
 \end{figure}
 \begin{figure}
 \includegraphics[angle=-90,width=0.45\textwidth]{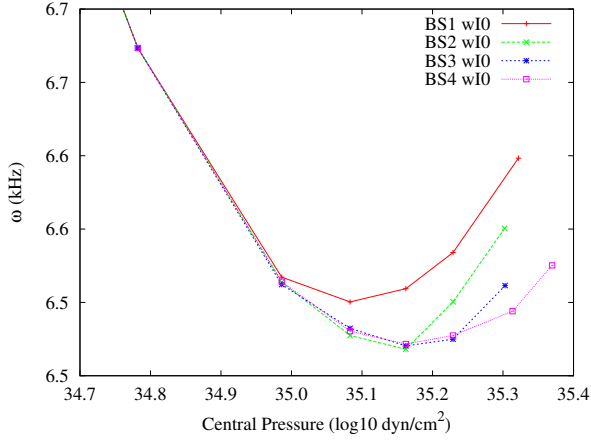} 
 \caption{Frequency of the fundamental wI mode vs central pressure for the BS
   equations of 
 state. The phase
   transitions that can be seen in the previous figure affects the frequency
   curves. These curves branch once the EOS is affected by the phase transition.}
 \label{eos_BS_wI0_real}
 \end{figure}
 \begin{figure}
 \includegraphics[angle=-90,width=0.45\textwidth]{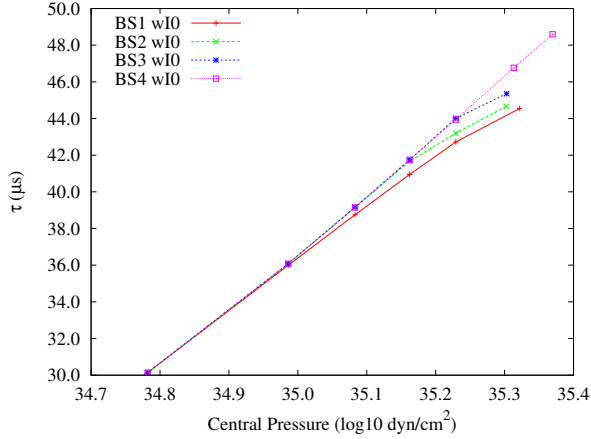} 
 \caption{Damping time of the fundamental wI mode vs central pressure for the
   BS equations of 
 state. In a
   similar way to the frequency, the phase transitions in the equations of
   state are clearly reflected in the branching of the damping time curves.}
 \label{eos_BS_wI0_imaginary}
 \end{figure}

\subsubsection{BS equations of state}
In order to study the influence of different phases in the compact star, we
study four equations of state, 
considered by Bonanno  and  Sedrakian with the NL3 parameterization of the
Walecka model \cite{Sedrakian}. These EOS correspond to hybrid stars with
hyperons and quark color-superconductivity. They usually have four regions,
which are, from the inside to the outside: CFL (quark matter three-flavor
color-flavor-locked), 
2SC (quark matter two-superconducting colors phase), N+Hyp (nuclear and
Hyper-nuclear matter), and N (nuclear matter). The extent of each
region depends on the quark-hadron transition density $\rho_{tr}$ and the
vector coupling constant $G_V$. We study four equations of state with $G_V =
0.6 G_S$ and $\rho_{tr}=2 \rho_0, 3 \rho_0, 3.5 \rho_0, 4\rho_0$ where
$\rho_0$ is the density of nuclear saturation.

In Figure \ref{eos_BS} we plot the equations of state in the high density
region. In Figure \ref{eos_BS_wI0_real} we plot the frequency of the 
fundamental wI 
mode as a function of the central pressure of the star in logarithmic scale.
In Figure \ref{eos_BS_wI0_imaginary}  
 we make a similar plot for the damping time of these
configurations. 

We can see in Figure \ref{eos_BS} that the BS equations of the state have
phase transitions, where for a given pressure, the  
density suffers a finite jump that increases its value (the EOS become
softer). The effect of these phase transitions is reflected clearly in the
frequency and  
damping time of the wI0 modes as it can be seen in Figures \ref{eos_BS_wI0_real}
and \ref{eos_BS_wI0_imaginary}. 

\section{Empirical relations for w-modes}
\begin{figure}
 \includegraphics[angle=-90,width=0.45\textwidth]{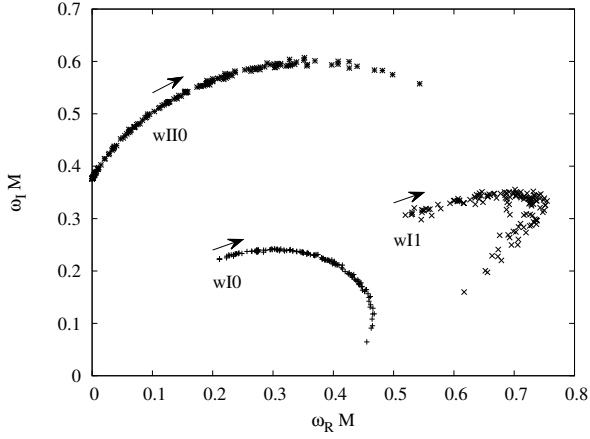} 
 \caption{Adimensional quantities $\omega_IM$ vs. $\omega_RM$ for all w-modes and EOS considered. Note that here we choose $c=G=1$, so the products $\omega_RM$, $\omega_IM$ are adimensional. It can be seen that the scaling with
   the mass is quite independent of the equation of state, specially for the
   fundamental modes. The arrow indicates increasing compactness.}
 \label{eos_w_modes_all_modes_KHZ_mus}
 \end{figure}

In Figure \ref{eos_w_modes_all_modes_KHZ_mus} we present $\omega_IM$
vs. $\omega_RM$ for wI0, wI1 and wII0 modes, for all
the equations of state studied. This is a plot similar to Fig.4 of
\cite{Andersson1996} for constant density stars. It can be seen that for the
fundamental modes, the scaling with the mass is quite independent of the
equation of state.  

\begin{figure}
 \includegraphics[angle=-90,width=0.45\textwidth]{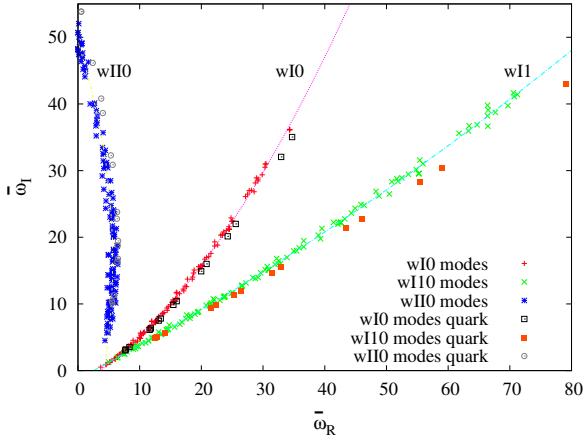} 
 \caption{$\bar\omega_{I}$ vs. $\bar\omega_{R}$ as defined in equations
   (\ref{empiricalpc}) for all the w-modes and for every equation of state we
   considered in the study. Here it can be seen that the scaling with 
   the central pressure is also quite independent of the equation of state. This
   universal relation could be used to estimate for example, the central
   pressure of the star independently of the EOS. The compactness decreases
   when increasing $\bar\omega_{I}$.}
 \label{eos_w_modes_all_modes_no_line}
 \end{figure}

Now let us define $\bar \omega_{R}$ and $\bar \omega_{I}$ (real and imaginary part of the w-modes in units of the central pressure) by the following relations: 
\begin{eqnarray}
\bar \omega_{R} &=&2 \pi  \frac{1}{\sqrt{p_c(cm^{-2})}}\frac{10^3}{c} \omega(Khz),\nonumber\\
\bar \omega_{I}&=& \frac{1}{\sqrt{p_c(cm^{-2})}}\frac{10^6}{c} \frac{1}{\tau(\mu
  s)} ,\label{empiricalpc}
\end{eqnarray}
where c is the speed of light in cm/s. If we plot $ \bar\omega_{R}$ versus  $\bar\omega_{I}$ of wI0, wI1, and wII0 modes for 
all the equations of state, we obtain Figure
\ref{eos_w_modes_all_modes_no_line} with three curves, one for each type of
modes. 
A remarkable feature is that all the wI0 modes of Figure
\ref{eos_w_modes_all_modes_KHZ_mus} can be described by the curve marked wI0 in Figure
\ref{eos_w_modes_all_modes_no_line},  almost independently of the  
 equation of state.  A similar behavior is obtained  for wI1 and wII0
 modes. One would expect all the wI0 modes of a given EOS to describe a
 curve depending on one parameter (for instance, the central pressure).
 And in principle, different equations of state should describe different
 curves. It is interesting to note that this is not the case, since the
 empirical relation between $ \bar\omega_{R}$ and $\bar\omega_{I}$ is
 independent of the equation of state. Hence all the modes of the same type
  are located in the same curve in Figure \ref{eos_w_modes_all_modes_no_line},
  independently of the EOS. Nevertheless, the parametrization of the
  curve using the central pressure depends on the EOS.
It is interesting to note that the only points that separate slightly  from the corresponding line are those of modes for equations of state WSPHS1 and WSPHS2 of quark stars.

This result allows us to  present an  empirical relation between $\bar \omega_{R}$
and $\bar \omega_{I}$  that seems to be independent of the
equation of the state. As far as 
we know this has not been obtained before.  

For all the EOS except WSPHS1 and WSPHS2, we obtain:

For the fundamental wI modes, our results of $\bar \omega_{I}$  versus  $\bar
\omega_{R}$ can be fitted to a quadratic relation as follows: 
\begin{eqnarray}
& & \bar\omega_{I}=(-2.30 \pm 0.13) + \label{frequencypcwI0}\\ & & (0.58 \pm 0.16)\bar\omega_{R} + 
(0.0165 \pm 0.0004)\bar\omega_{R}^{2},   \nonumber
\end{eqnarray}
with $\chi^{2}=0.107$.

For the first excited wI mode, we obtain the following fit:
\begin{eqnarray}
& & \bar\omega_{I}=(-1.22 \pm 0.13) + \\ & & (0.485 \pm 0.008)\bar\omega_{R} +
(0.00163 \pm 0.00011)\bar\omega_{R}^{2}, \nonumber
\end{eqnarray}
with $\chi^{2}=0.170$. It is almost linear with $\bar
\omega_{R}$.

For the fundamental wII mode, $\bar
\omega_{R}$ is quadratic with $\bar
\omega_{I}$. We obtain the following fit:
\begin{eqnarray}
& & \bar\omega_{R}=(4.21 \pm 0.14) + \\& &  (0.173 \pm 0.012)\bar\omega_{I} +
(-0.00524 \pm 0.00020)\bar\omega_{I}^{2},  \nonumber
\end{eqnarray}
with $\chi^{2}=0.165$.

For quark stars (EOS WSPHS1 and WSPHS2) we obtain slightly different values for the parameters 
of the empirical relation:

For the fundamental wI modes, our results of $\bar
\omega_{I}$  versus $\bar
\omega_{R}$ can be fitted to a quadratic relation as follows:
\begin{eqnarray}
& & \bar\omega_{I}=(-2.20 \pm 0.22) + \label{frequencypc}\\ & & (0.557 \pm 0.024)\bar\omega_{R} + 
(0.0148 \pm 0.0006)\bar\omega_{R}^{2},   \nonumber
\end{eqnarray}
with $\chi^{2}=0.0220$.

For the first excited wI mode, we obtain the following fit:
\begin{eqnarray}
& & \bar\omega_{I}=(-1.275 \pm 0.097) + \\ & & (0.476 \pm 0.005)\bar\omega_{R} +
(0.0010 \pm 0.00005)\bar\omega_{R}^{2}, \nonumber
\end{eqnarray}
with $\chi^{2}=0.00902$. It is almost linear with $\bar
\omega_{R}$.

For the fundamental wII mode, $\bar
\omega_{R}$ is quadratic with $\bar
\omega_{I}$. We obtain the following fit:
\begin{eqnarray}
& & \bar\omega_{R}=(4.32 \pm 0.30) + \\& &  (0.198 \pm 0.022)\bar\omega_{I} +
(-0.0051 \pm 0.0004)\bar\omega_{I}^{2},  \nonumber
\end{eqnarray}
with $\chi^{2}=0.061$. 

These relations allow us to obtain $\bar\omega_{I}$ if
$\bar\omega_{R}$ is known. From a practical point of view, these relations can
be used to simplify the numerical calculations to obtain quasi-normal modes
for any equation of state, reducing, in our method, the time necessary  
to find the zeros of the determinant.

Let us present  a possible application of these empirical relations. Suppose that the frequency $\omega(Khz)$ and 
the damping time $\tau(\mu s)$ of the fundamental mode wI0 of a neutron star are detected. Then, using the equation (\ref{empiricalpc})  we can plot a line for 
$\bar\omega_{I}$ versus $ \bar\omega_{R}$ in Figure
\ref{eos_w_modes_all_modes_no_line} with parameter of the line $p_c$. The
crossing point of this line with the curve marked wI0, given by the empirical
relation (\ref{frequencypcwI0}), gives us an estimation of the central pressure
$p_c$ independently of the equation of state. Now, we can check which equations
of state are compatible with this $p_c$, i.e., which EOS  
have the measured wI0 mode near the crossing point for the estimated central
pressure. Hence, this method allows to estimate the central pressure $p_c$
and, in some situations, discard certain equations of state. Let us mention,
that using this $p_c$, we can estimate  
the mass and the radius of the star using our calculations for the possible
EOS. Note, that if 
these data (mass and radius) are measured by other methods, we would have another filter
to impose to the EOS.
\section{Other results}

Finally,  we  study the effect of core-\textit{crust} transition
pressure variations in the quasi-normal mode spectrum using our model of crust
as a 
surface energy density enveloping the 
core. We will consider the SLy equation
of state, for which the core-\textit{crust} transition
is found at density $1.29\cdot10^{14} g/cm^{3}$. We construct a star of
$1.4M_{\odot}$ with and without a surface energy density enveloping the core
of the star at this core-\textit{crust} transition region. We can compare the difference
in the fundamental w modes between our approximation of \textit{crust} as a thin
shell enveloping the core, and the standard calculation. The result is the
following: for the fundamental wI mode, the change in the frequency is
$0.07\%$ and the variation in the damping time is $-0.02\%$. For the
fundamental wII mode, the changes are bigger. For the frequency $0.3\%$ and
for the damping time $-0.07\%$. This indicates that this approximation makes small changes in
the w quasi-normal spectrum. 

Next, we consider this neutron star with $1.4M_{\odot}$ and surface
energy 
density at the core-\textit{crust} transition point: $4.8\cdot10^{32}
dyn/cm^{2}$. Fixing the central pressure, we variate the core-\textit{crust} transition
pressure between $10^{32}
dyn/cm^{2}$ and $10^{33}
dyn/cm^{2}$, and compare with the modes for the initial configuration. 
The
impact of transition pressure variations on the frequency is small, between
$-0.05\%$ and 
$0.08\%$ for the fundamental wI mode, and between $-0.24\%$ and
$0.35\%$ for the fundamental wII mode. 
Similarly, transition pressure variations has a
small impact on the damping time, between $0.02\%$ and
$-0.035\%$ for the fundamental wI mode, and between $0.06\%$ and
$-0.1\%$ for the fundamental wII mode.
\section{Summary}
In this paper we have studied the axial w quasi-normal modes for 18 equations
of state. In particular, we consider the influence of the presence of exotic matter
in the core of the neutron star. 

We have seen that the hyperon core changes the frequency and the damping time
of the modes for configurations of high compactness. The frequency increases
with the central pressure above $\sim 10^{35} dyn/cm^2$, where the hyperon
phases are more important due 
to the high density of the star core. This effect is related to
the softening of the equation of state  at the inner regions of the
star. For stars including quark matter in the core, we obtain  a
clear influence of the quark presence  on the frequency 
and damping time when represented versus central pressure, with a
characteristic behavior for all the EOS with quark matter in the core.
 
  We have obtained phenomenological relations for the frequency and damping
  time with $M/R$ for wI and wII modes. For some equations of state (for
  hyperon matter H1 and BGN1H1, and for pure quark stars WSPHS1-2) the scaling
  relations are quite different from the rest of EOS. For the rest of
  EOS considered, the scaled relations are quite similar,
  providing useful information that  could be used to measure the radius of a
  neutron star.  

A detailed analysis of four equations of state
considered by Bonanno  and  Sedrakian allows us to study the influence of the
presence of different phase transitions in the spectrum of a compact star. 

 Another interesting result is the new phenomenological relation between the
real part $\bar
\omega_{R}$ and the imaginary part $\bar
\omega_{I}$ (scaled to the central pressure) of w quasi-normal modes, valid
 for all the EOS, except, perhaps, for WSPHS1 and WSPHS2 EOS (quark stars).  

The precision of our algorithm allows us to construct explicitly the universal
low compactness limiting configuration around $M/R=0.106$ for which the
fundamental wII mode vanishes. The existence of this configuration was
conjectured by \cite{Wen2009}.  

Finally, we have studied the influence of  changes in the core-\textit{crust}
transition pressure obtaining that the influence is very small. 

In order to perform this analysis, we developed a new numerical method to
calculate w-axial quasi-normal-modes. The use of the phase of the Regge-Wheeler
function together with Exterior Complex Scaling, allow us to impose the
outgoing wave behavior with stringent conditions. We allow for variable 
angle in the complex scaling, which makes it possible
to obtain new modes and
enhances precision with respect previous works.

We present a complete study of the junction conditions between the exterior
and interior 
solution in a surface of constant
pressure (also including the case $p=0$, the border of the star).
The matching conditions can be written in terms of a determinant whose zeros
correspond 
to quasi-normal modes for
the particular integrated configuration.
\begin{acknowledgements}
We would like to thank I. Bednarek for kindly providing
us with the EOS BHZBM, I. Sagert for EOS WSPHS1-3, A. Sedrakian for EOS BS1-4, 
and S. Weissenborn for EOS WCS1-2.  
We thank Prof. Kostas D. Kokkotas for very helpful discussions concerning
quasi-normal modes of neutron stars. J.L.B wants to thank also his kind
hospitality. We are very grateful to Prof. Gabriel A. Galindo for his
help concerning the Exterior Complex Scaling method. We also want to thank the referee
for his/her very helpful comments and suggestions. 
 This work was supported by the Spanish Ministerio de Ciencia e Innovacion, 
 research project FIS2011-28013. J.L.B was supported by the Spanish 
 Universidad Complutense de Madrid.  
\end{acknowledgements}

\bibliography{axial_bib}

\end{document}